%% file: mltipreclscape.tex
\documentclass{article}
\pdfoutput=1 
\usepackage{wrapfig}
\usepackage{subfig}
\usepackage{array}
\usepackage{hyperref}
\usepackage{multicol}
\usepackage{listings}
\usepackage{xcolor}
\newcolumntype{L}[1]{>{\raggedright\let\newline\\\arraybackslash\hspace{0pt}}m{#1}}
\definecolor{Gray}{gray}{0.85}

\definecolor{Gray}{gray}{0.85}
\definecolor{LightCyan}{rgb}{0.88,1,1}
\definecolor{LightGreen}{rgb}{0.67, 0.88, 0.69}
\definecolor{LightBlue}{rgb}{0.6, 0.73, 0.89}
\definecolor{LightYellow}{rgb}{0.98, 0.91, 0.71}
\definecolor{LightRed}{rgb}{0.98, 0.81, 0.69}
\usepackage{kpfonts}

\usepackage{amstext,amsmath,amssymb,amsfonts,amsthm}

\newcolumntype{g}{>{\columncolor{Gray}}c}
\newcolumntype{w}{>{\columncolor{white}}c}
\newcolumntype{y}{>{\columncolor{LightCyan}}c}

\usepackage[framemethod=TikZ,backgroundcolor=gray!40,shadow=true,roundcorner=8pt]{mdframed}
\usetikzlibrary{shadows}
\usepackage{hyperref}
\hypersetup{
	colorlinks,
	citecolor=blue,
	filecolor=blue,
	linkcolor=blue,
	urlcolor=blue
}

\usepackage{algorithm}
\usepackage{algpseudocode}
\usepackage{cleveref}

\usepackage{cellspace}
\usepackage{multirow}

\iftrue
\newcommand{\team}[1] {\textcolor{cyan}{[#1]}}
\else
\newcommand{\team}[1] {}
\fi

\newcommand{\eps}{\varepsilon}

\usepackage{acro}
\DeclareAcronym{asic}{short=ASIC,long=Application-Specific Integrated Circuit}
\DeclareAcronym{blas}{short=BLAS,long=Basic Linear Algebra Subprograms}
\DeclareAcronym{fpga}{short=FPGA,long=Field Programmable Gate Array}
\DeclareAcronym{gmres}{short=GMRES,long=Generalized Minimum Residual}
\DeclareAcronym{hpc}{short=HPC,long=high-performance computing}

\newcommand{\fpprec}[2]{#1^{(#2)}}
\usepackage[preprint]{neurips_2020}

\author{by the ECP Multiprecision Effort Team (Lead: Hartwig Anzt)\\
  \\
  \begin{minipage}{\columnwidth}
    \centering
    Ahmad Abdelfattah\textsuperscript{1}, 
    Hartwig Anzt\textsuperscript{1,2}, 
    Erik G. Boman\textsuperscript{3},
    Erin Carson\textsuperscript{4}, 
    Terry Cojean\textsuperscript{2},
    Jack Dongarra\textsuperscript{1,5,6},
    Mark Gates\textsuperscript{1},
    Thomas Gr\"{u}tzmacher\textsuperscript{2}, 
    Nicholas J. Higham\textsuperscript{6}, 
    Sherry Li\textsuperscript{8}, 
    Neil Lindquist\textsuperscript{1},
    Yang Liu\textsuperscript{8}, 
    Jennifer Loe\textsuperscript{3}, 
    Piotr Luszczek\textsuperscript{1}, 
    Pratik Nayak\textsuperscript{2},
    Sri Pranesh\textsuperscript{6},
    Siva Rajamanickam\textsuperscript{3},
    Tobias Ribizel\textsuperscript{2}, 
    Barry Smith\textsuperscript{9},
    Kasia Swirydowicz\textsuperscript{10},
    Stephen Thomas\textsuperscript{10},
    Stanimire Tomov\textsuperscript{1}, 
    Yaohung M. Tsai\textsuperscript{1}, 
    Ichi Yamazaki\textsuperscript{3},
    Urike Meier Yang\textsuperscript{7}\\
    \vspace{1cm}
    \textsuperscript{1}University of Tennessee, Knoxville, USA\\
    \textsuperscript{2}Karlsruhe Institute of Technology, Karlsruhe, Germany\\
    \textsuperscript{3}Sandia National Lab, Albuquerque, USA\\
    \textsuperscript{4}Charles University, Prague, Czech Republic\\
    \textsuperscript{5}Oak Ridge National Lab, Oak Ridge, USA\\
    \textsuperscript{6}University of Manchester, Manchester, UK\\
    \textsuperscript{7}Lawrence Livermore National Lab, USA\\
    \textsuperscript{8}Lawrence Berkeley National Lab, Berkeley, USA\\
    \textsuperscript{9}Argonne National Lab, Argonne, USA\\
    \textsuperscript{10} National Renewable Energy Lab, Boulder, USA
\end{minipage}
}

\title{A Survey of Numerical Methods Utilizing Mixed Precision Arithmetic}
\date{\today}

\begin{document}
\maketitle


%





\clearpage
\input{intro}
\input{densela}

\input{compression}
\input{sparse_factorization}

\input{krylov}
\input{preconditioning}
\input{format_decoupling}

\input{multigrid}
\input{fft}
\input{ml}

\input{interoperability}
\input{formats}

\section*{Acknowledgments}
This work was supported by the US Exascale Computing Project 
(17-SC-20-SC), a collaborative effort of the U.S. Department of Energy Office 
of Science and the National Nuclear Security Administration.
This work was performed under the auspices of the U.S. Department of Energy by Lawrence Livermore National Laboratory under Contract DE-AC52-07NA27344.

\section*{Disclaimer}
This document was prepared as an account of work sponsored by an agency of the United States government. Neither the United States government nor Lawrence Livermore National Security, LLC, nor any of their employees makes any warranty, expressed or implied, or assumes any legal liability or responsibility for the accuracy, completeness, or usefulness of any information, apparatus, product, or process disclosed, or represents that its use would not infringe privately owned rights. Reference herein to any specific commercial product, process, or service by trade name, trademark, manufacturer, or otherwise does not necessarily constitute or imply its endorsement, recommendation, or favoring by the United States government or Lawrence Livermore National Security, LLC. The views and opinions of authors expressed herein do not necessarily state or reflect those of the United States government or Lawrence Livermore National Security, LLC, and shall not be used for advertising or product endorsement purposes.\\

Sandia National Laboratories is a multimission laboratory managed and operated by National Technology and Engineering Solutions of Sandia, LLC., a wholly owned subsidiary of Honeywell International, Inc., for the U.S. Department of Energy’s National Nuclear Security Administration under contract DE-NA-0003525.

\bibliographystyle{unsrt}
\clearpage
\bibliography{references}

\end{document}

%% file: intro.tex
\section{Introduction}
Within the past years, hardware vendors have started designing low precision special function units 
in response to the demand of the Machine Learning community and their demand for high compute power
in low precision formats. Also the server-line products are increasingly featuring low-precision
special function units, such as the NVIDIA tensor cores
in ORNL's Summit supercomputer providing more than an order of magnitude higher performance
than what is available in IEEE double precision. At the same time, the gap between the compute power
on the one hand and the memory bandwidth on the other hand keeps increasing, 
making data access and communication prohibitively expensive compared to arithmetic operations.
Having the choice between ignoring the hardware trends and continuing the traditional path, 
and adjusting the software stack to the changing hardware designs, the US Exascale Computing Project
decided for the aggressive step of building a multiprecision focus effort to take on the 
challenge of designing and engineering novel algorithms exploiting the compute power 
available in low precision and adjusting the communication format to the application specific needs.
To start the multiprecision focus effort, we survey the numerical linear algebra community 
and summarize all existing multiprecision knowledge, expertise, and software capabilities
in this landscape analysis report. We also include current efforts and preliminary results that 
may not yet be considered ```mature technology,'' but have the potential to grow into
production quality within the multiprecision focus effort.
As we expect the reader to be familiar with the basics of numerical linear algebra,
we refrain from providing a detailed background on the algorithms themselves but focus on how
mixed- and multiprecision technology can help improving the performance of these methods 
and present highlights of application significantly outperforming the traditional fixed precision 
methods.

\clearpage

\tableofcontents

\clearpage

%% file: densela.tex
\section{Dense Linear Algebra}
\input{lowp_blas}

\subsection{Classical Iterative Refinement}
\label{sec:denseiterref}
\label{sec:}

On modern architectures, the performance of 32-bit operations is often at least twice as fast as the performance of 64-bit operations. There are two reasons for this. Firstly, 32-bit floating point arithmetic is usually twice as fast as 64-bit floating point arithmetic on most modern processors. Secondly the number of bytes moved through the memory system is halved.

Mixed precision algorithms stem from the observation that, in many cases, a single precision solution of a problem can be refined to the point where double precision accuracy is achieved. The refinement can be accomplished, for instance, by means of the Newton’s algorithm (see \Cref{eq:newton}) which computes the zero of a function $f(x)$ according to the iterative formula

\begin{equation}
    \label{eq:newton}
    x_{n+1} = x_n - \frac{f(x_n)}{f'(x_n)}
\end{equation}

In general, we would compute a starting point and $f(x)$ in single precision arithmetic and the refinement process will be computed in double precision arithmetic.

If the refinement process is cheaper than the initial computation of the solution, then double precision accuracy can be achieved nearly at the same speed as the single precision accuracy.

A common approach to the solution of linear systems, either dense or sparse, is to perform the LU factorization of the coefficient matrix using Gaussian elimination. First, the coefficient matrix $A$ is factored into the product of a lower triangular matrix $L$ and an upper triangular matrix $U$. Partial row pivoting is in general used to improve numerical stability resulting in a factorization $PA = LU$, where $P$ is a permutation matrix. The solution for the system is achieved by first solving $Ly = Pb$ (\textit{forward substitution}) and then solving $Ux = y$ (\textit{backward substitution}). Due to round-off errors, the computed solution $x$ carries a numerical error magnified by the condition number of the coefficient matrix $A$.

In order to improve the computed solution, we can apply an iterative process which produces a correction to the computed solution at each iteration, which then yields the method that is commonly known as the \textit{iterative refinement} (IR) algorithm. As Demmel points out \cite{demmelapplied}, the nonlinearity of the round-off errors makes the iterative refinement process equivalent to the Newton’s method applied to the function $f(x) = b - Ax$. Provided that the system is not too ill-conditioned, the algorithm produces a solution correct to the working precision. Iterative refinement in double/double precision is a fairly well understood concept and was analyzed by Wilkinson \cite{wilkinson1994rounding}, Moler \cite{moler1967iterative} and Stewart \cite{stewart1973introduction}.

Iterative refinement is a long-standing method that was programmed by Wilkinson in the 1940s for the ACE digital computer. The idea is to improve the computed solution of a linear system by iteratively solving a correction equation and adding the correction to the current solution; for a comprehensive treatment, Higham \cite[Chap.~12]{higham2002accuracy}.

The algorithm can be modified to use a mixed precision approach. The three tasks— original solve/factorization, residual computation, and correction equation solve—can be done in the same precision (fixed precision) or in different precisions (mixed precision). The original usage was mixed precision, with the residual computed at twice the working precision. 

For the current work, the factorization $PA = LU$ and the solution of the triangular systems $Ly = Pb$ and $Ux = y$ are computed using single precision arithmetic. The residual calculation and the update of the solution are computed using double precision arithmetic and the original double precision coefficients (see \Cref{alg:mix_precision_IR_DS}). The most computationally expensive operation, the factorization of the coefficient matrix $A$, is performed using single precision arithmetic and takes advantage of its higher speed. The only operations that must be executed in double precision are the residual calculation and the update of the solution (they are denoted with an $\epsilon_d$ in \Cref{alg:mix_precision_IR_DS}). 

\begin{algorithm}
    \caption{Mixed precision, Iterative Refinement for Direct Solvers}
    \label{alg:mix_precision_IR_DS}
    \begin{algorithmic}[1]
    \State $LU \gets PA$ \Comment{($\epsilon_s$)}
    \State Solve $Ly = Pb$ \Comment{($\epsilon_s$)}
    \State Solve $Ux_0 = y$ \Comment{($\epsilon_s$)}
    \For{$k = 1, 2, \dots$}
        \State $r_k \gets b - Ax_{k-1}$ \Comment{($\epsilon_d$)}
        \State Solve $Ly = Pr_k$\label{alg:mix_precision_IR_DS:loop_L} \Comment{($\epsilon_s$)}
        \State Solve $Uz_k = y$\label{alg:mix_precision_IR_DS:loop_U} \Comment{($\epsilon_s$)}
        \State $x_k \gets x_{k-1} + z_k$ \Comment{($\epsilon_d$)}
        \State Check convergence
    \EndFor
    \end{algorithmic}
\end{algorithm}

We observe that the only operation with computational complexity of $O(n^3)$ is handled in single precision, while all operations performed in double precision are of at most $O(n^2)$ complexity. The coefficient matrix $A$ is converted to single precision for the LU factorization and the resulting factors are stored in single precision while the initial coefficient matrix $A$ needs to be kept in memory. Therefore, one drawback of the following approach is that the it uses 50\% more memory than the standard double precision algorithm.

The method in \Cref{alg:mix_precision_IR_DS} can offer significant improvements for the solution of a sparse linear system in many cases if:
\begin{enumerate}
    \item single precision computation is significantly faster than double precision computation;
    \item the iterative refinement procedure converges in a small number of steps;
    \item the cost of each iteration is small compared to the cost of the system factorization. If the cost of each iteration is too high, then a low number of iterations will result in a performance loss with respect to the full double precision solver. In the sparse case, for a fixed matrix size, both the cost of the system factorization and the cost of the iterative refinement step may substantially vary depending on the number of nonzeroes and the matrix sparsity structure; this will be addressed in Section~\ref{sec:sparse-IR}. In the dense case, results are more predictable.
\end{enumerate}
Note that the choice of the stopping criterion in the iterative refinement process is critical. Formulas for the errors computed at each step of \Cref{alg:mix_precision_IR_DS} can be obtained for instance in \cite{demmel2006error,oettli1964compatibility}.

Recently, Carson and Higham \cite{carson2018accelerating} analyzed a three-precision iterative refinement scheme (factorization precision, working precision, residual precision) and concluded that if the condition number of A is not too large, namely $\kappa_\infty(A)=\left\lVert A \right\rVert_\infty \left\lVert A^{-1} \right\rVert_\infty < 10^4$, then using FP16 for the $O(n^3)$ portion (the LU factorization) and (FP32, FP64) or (FP64, FP128) as the (working, residual) precision for the $O(n^2)$ portion (refinement loop), one can expect to achieve forward error and backward error on the order of $10^{-8}$ and $10^{-16}$ respectively. We note that, if $\hat{x}$ is an approximate solution of $Ax = b$ the forward error is defined by $\frac{ \left\lVert\hat{x}-x \right\rVert_\infty}{\left\lVert x \right\rVert_\infty}$ and the backward error is defined by $\min\{\epsilon: (A + \Delta A) \hat{x} = b, \left\lVert \Delta A \right\rVert \leq \epsilon\left\lVert A \right\rVert\}$ and can be evaluated as $\frac{\left\lVert r \right\rVert_2}{\left\lVert A \right\rVert_2 \left\lVert \hat{x} \right\rVert_2}$, where $r = b-A\hat{x}$. 

\subsection{GMRES-IR}


Carson and Higham \cite{carson2017new} proposed the use of \ac{gmres}
\cite{saad1986gmres} preconditioned by the FP16 LU factors 
as the solver in correction equation
and showed that in this case the constraint on the condition number 
can be relaxed to $\kappa_\infty(A) < 10^8$ when the (working, residual) 
precision is (FP32, FP64) and to $10^{12}$ when the (working, residual) 
precision is (FP64, FP128). We refer to \cite[Table 2.2]{high19i}
for limiting condition number, and forward, backward errors.
Analysis covering this \ac{gmres}-based 
approach when two precisions are used with the residual precision 
equal to the working precision is given in \cite{high19i}.

 The idea is that the \ac{gmres} solver will 
provide a better and more stable approximate solution to $Az_k=r_k$ 
than the basic triangular solve, which is directly affected by the 
quality of the low-precision LU factors. Using \ac{gmres}, we can still 
guarantee that the solution of the correction equation $Az_k=r_k$ 
has some correct digits and a residual at the level of the 
convergence tolerance requested by the algorithm. The convergence 
tolerance of the refinement process is chosen to be of the order of 
the unit roundoff of 
the low-precision arithmetic used during the factorization
(e.g., we use $10^{-4}$ or $10^{-8}$ when the LU factorization is in 
FP16 or FP32, respectively). This variant is called \ac{gmres}-based
iterative refinement (GMRES-IR) by Carson 
and Higham, and it is described in Algorithm~\ref{alg:IRGM}. Note that $U^{-1}$ 
and $L^{-1}$ are never explicitly formed; instead matrix–vector products 
$U^{-1}L^{-1}Ay$ needed by \ac{gmres} are computed by multiplication by $A$ 
followed by two triangular solves. 
Since this paper focuses on the practical usage and possible performance gains rather than error analysis, we point the reader to \cite{higham2002accuracy,carson2017new,high19i} for detailed error analysis of the IR and GMRES-IR techniques.

\begin{algorithm}
  \caption{Iterative refinement using \ac{gmres} (GMRES-IR)}
    \label{alg:IRGM}
    \begin{algorithmic}[1]
    \State Convert $A$ to $A_f$ from precision $u_w$ to $u_f$
    \State Perform LU factorization of $A_f$ in precision $u_f$
    \State Find the initial solution $x_0$ using the computed LU factorization of $A_f$ in precision $u_f$ then cast $x_0$ to precision $u_w$ 
    \State // Refinement loop, outer loop
    \Repeat
        \State Compute residual $r_k = b - Ax_k$ in precision $u_r$ and cast it to $u_w$ \Comment{Residual}
        \State Solve $U^{-1}L^{-1}Az_k = U^{-1}L^{-1}r_k$ by \ac{gmres} in precision $u_w$ \Comment{Correction}
        \State Correct the current solution $x_{k+1} = x_k + z_k$ in precision $u_w$ \Comment{Update}
    \Until{$x_k$ is accurate enough}
    \end{algorithmic}
\end{algorithm}

The design and implementation of numerical algorithms that efficiently exploit 
current highly parallel computer architectures is a challenge, especially 
if close to peak performance is to be achieved. Since in the GMRES-IR approach 
$\mathcal{O}(n^3)$ operations are in lower precision, this idea allows a new class of iterative refinement solvers and a number of computational techniques that allow us to solve fundamental $Ax = b$ problems close to peak FP64 performance. The developments open up directions for future work, including further optimizations, development of a full set of mixed-precision factorizations, linear system solvers as well as eigensolvers and singular value decomposition (SVD).

\subsubsection{Scaling}
It is clear that the use of low precision floating-point 
arithmetic in iterative refinement 
can lead to significant speedups. However, fp16 has a small dynamic
range, and therefore encountering overflow, underflow, and
subnormal numbers is very likely. To address these issues 
now we discuss scaling algorithms, which are presented in \cite{hpz19}, \cite{hipr19a}
and \cite{chp20}. We refer interested readers to these references
for more details.

We consider a two-sided diagonal scaling
prior to converting to fp16: $A$ is replaced by $RAS$, where 
$$
       R = \text{diag}(r_i), \quad S = \text{diag}(s_i), \qquad 
         r_i, s_i >0,  \quad i=1\colon n.
$$
Such scaling algorithms have been developed in the 
context of linear systems and linear programming problems.  
Despite the large literature on scaling such problems,
no clear conclusions are available on when or how one should scale;
see \cite{elsa12} for a recent experimental study.
In contrast to previous studies, where the aim of scaling has been to reduce a condition number or to speed up the convergence of an iterative method applied to the scaled matrix, we scale in order to help squeeze a single or double precision matrix into half precision, with a particular application to using the resulting half precision LU factors for iterative refinement.

Our usage of two-sided diagonal scaling is given in
Algorithm~\ref{alg.apply-diag-scale}.

\begin{algorithm}
\caption{(Two-sided diagonal scaling then round). This algorithm rounds 
$A\in\mathbb{R}^{n \times n}$ 
to the fp16 matrix $A^{(h)}$, 
scaling all elements to avoid overflow. $\theta\in(0,1]$ is a parameter.}
\label{alg.apply-diag-scale}
\begin{algorithmic}[1]
\State Apply any two-sided diagonal scaling algorithm to $A$, to obtain diagonal
matrices $R$, $S$.
\State Let $\beta$ be the maximum magnitude of any entry of $RAS$.
\label{line.beta}%
\State $\mu = \theta x_{\text{max}} /\beta$
\State $A^{(h)} = fl_h( \mu (RAS))$ 
\end{algorithmic}
\end{algorithm} 

We consider two different algorithms for determining $R$ and $S$;
both algorithms are carried out at the working precision.
One option is row and column equilibration, which ensures that every row and
column has maximum element in modulus equal to $1$---that is, each row and
column is equilibrated. The LAPACK routines \texttt{xyyEQU} carry out this form of 
scaling~\cite{lug99}.
A symmetry-preserving two-sided scaling is proposed by Knight, Ruiz, and U{\c{c}}ar~\cite{kru14}. The algorithm is iterative and scales simultaneously on both sides rather than sequentially on one side then the other. 

\subsection{Mixed-precision Factorizations}

Haidar at al.~\cite{haidar2018harnessing} proposed IR methods using mixed-precision factorizations. While classical IR and extensions like the GMRES-IR use fixed-precision factorizations (e.g., in precision $u_f$ as illustrated in Algorithm~\ref{alg:IRGM}), mixed-precision factorizations apply higher precision (e.g., $u_w$) at critical parts of the algorithm to get extra-precise factorizations while retaining the performance of the low-precision counterpart. The developments were applied to GPU Tensor Cores and illustrate that FP16 can be used to get FP64 accuracy for problems with $\kappa_\infty(A)$ of up to $10^5$, compared to a more typical requirement of $\kappa_\infty(A) < 10^4$. The work illustrates that mixed-precision techniques can be of great interest for linear solvers in many engineering areas. The results show that on single NVIDIA V100 GPU the new solvers can be up to four times faster than an optimized 
double precision solver \cite{haidar2018harnessing},\cite{haidar2017investigating}, \cite{hazw18}.

The mixed-precision factorizations were motivated by the need to get extra precision when working with very low precisions, like the FP16. Also, this allows to easily overcome implementation issues and other limitations of using FP16 arithmetic, and thus harness the power 
of specialized hardware, like the GPU Tensor Cores, for a larger 
range of scientific computing applications. 

A building-block for the mixed-precision factorizations is mixed-precision BLAS.
Having mixed-precision BLAS can enable 
the ease of developing many mixed-precision LAPACK algorithms.
Currently, cuBLAS provides mixed FP32-FP16 precision HGEMM
that uses the GPU's Tensor Cores FP16 acceleration. 
In this GEMM, the input {\tt A} and {\tt B} matrices can be FP32, 
internally get casted to FP16, used to compute a GEMM on Tensor 
Cores in full (FP32) accuracy, and the result stored back on the
GPU memory in FP32. There are two main benefits of having such
mixed-precision BLAS routines. First, note that this mixed-precision
HGEMM is almost as fast as the non-mixed FP16 precision only HGEMM (see Figure~\ref{fig:hgemm_performance}), and second, 
the use of mixed-precision gains about one more decimal digit of 
accuracy (see Figure~\ref{fig:hgemm_error}).

\begin{figure}[tb]
\centering
\includegraphics[width=0.7\linewidth]{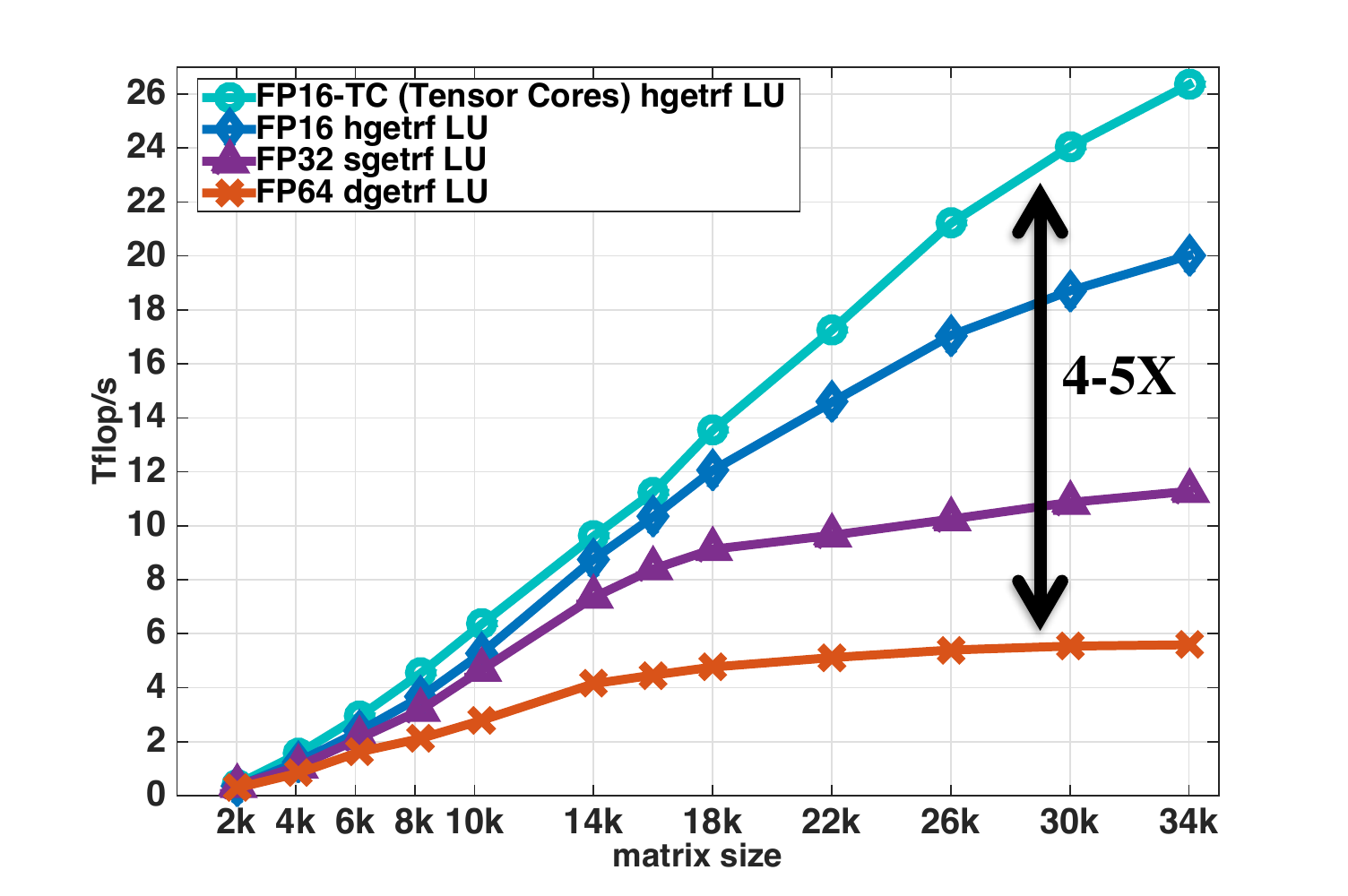}
\caption{Mixed-precision LU (hgetrf) in MAGMA and its 
         speedup vs. FP64 LU.}
\label{fig:fused}
\end{figure}

Besides the two main benefits outlined above, the availability 
of mixed-precision GEMM also enables us
to easily develop other mixed-precision algorithms, e.g., 
LAPACK, and in particular, the various mixed-precision 
factorizations that we recently added in 
MAGMA~\cite{haidar2018harnessing}. Figure~\ref{fig:fused}
shows the performance of the mixed-precision LU (marked as
"FP16-TC hgetrf LU"). Note that this factorization is about 
$4-5\times$ faster than dgetrf. Its data storage is in FP32
and the implementation is the same as sgetrf, except that it uses
the mixed-precision HGEMMs for the trailing matrix updates.

\begin{figure}[tb]
\centering
\includegraphics[width=0.7\linewidth]{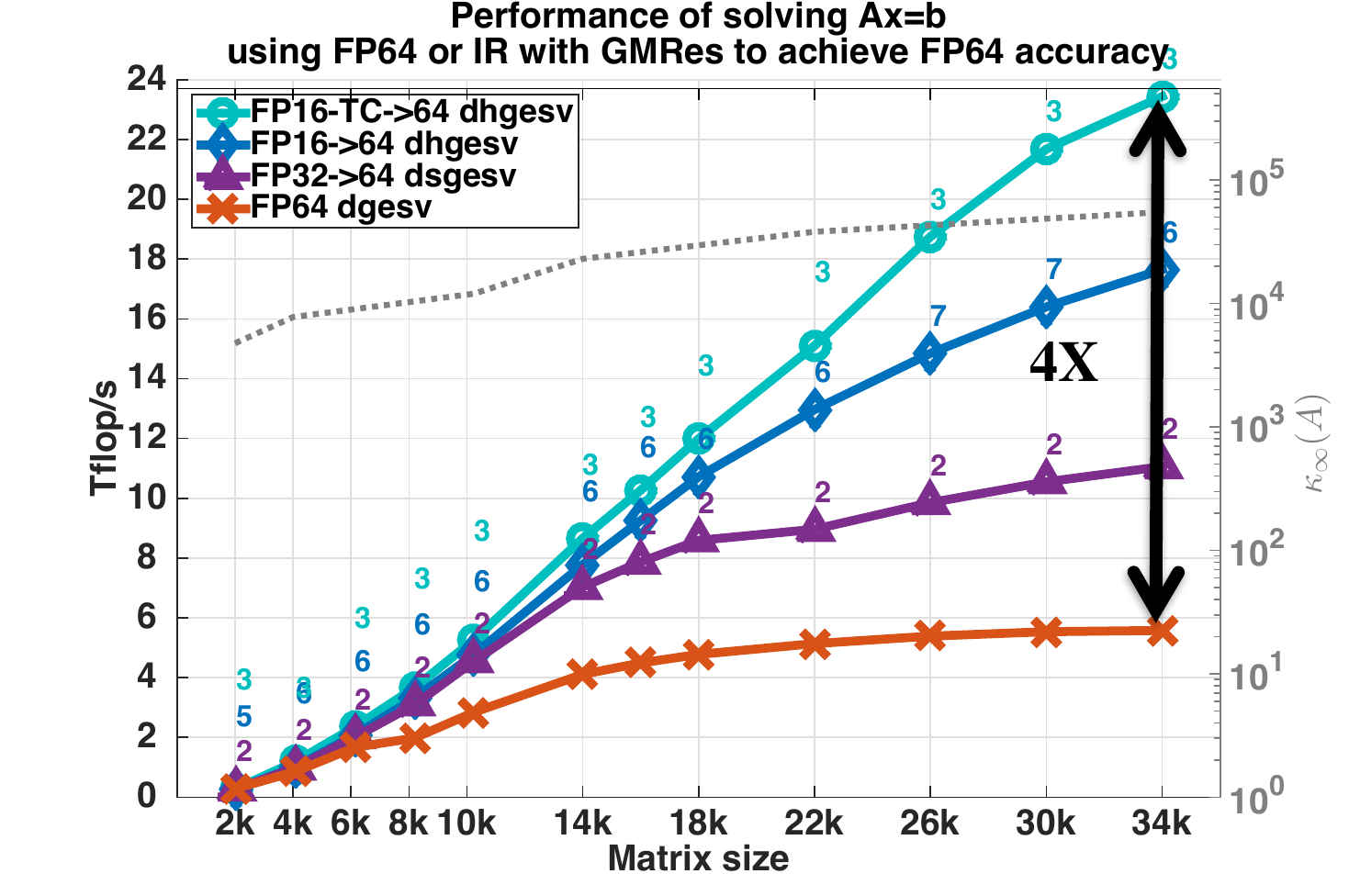}
\caption{Mixed-precision 
         iterative refinement in MAGMA and acceleration vs.
         FP64 solvers. Note $\approx2\%$ overhead per iteration,
         and more than $2\times$ less overhead in terms of iterations
         for mixed-precision LU vs. regular FP16 LU (the 3 vs. 7 
         iterations until FP64 convergence).}
\label{fig:fusedRight}
\end{figure}

Figure~\ref{fig:fusedRight} shows the mixed-precision iterative 
refinement in MAGMA~\cite{haidar2018harnessing}. The $4\times$ overall acceleration is due to a number of optimizations. First, note
that the 3 iterations to get to FP64 accuracy led to loss of about 
2 Tflop/s compared to the hgetrf performance (24 Tflop/s vs. 26 Tflop/s), i.e., the overhead of one iteration can be deduced as
being about 2\%.
Loosing 75\%, e.g., through up to 40 iterations, would lead to
no acceleration compared to FP64 solver. This overhead per iteration is very 
low, which is due to fusing all data conversions with computational
kernels. Without fusion, the overhead would have been easily about 
$3\times$ higher. Second, note that the iterative refinement 
using the mixed-precision factorization has more than $2\times$
smaller overhead in terms of iterations to solution (the $3$ vs. $7$
iterations until FP64 convergence). This is due to the extra digit 
of accuracy that the mixed-precision HGEMM has over the FP16 HGEMM,
which also translates to a more accurate mixed-precision LU.

\subsection{Cholesky Factorization} \label{sec.lp-chol}


In the previous section we considered scaling of a general and 
symmetric matrix, we now assume that we are given a symmetric positive definite matrix 
$A\in\mathbb{R}^{n \times n}$
in finite precision arithmetic of precision $u$
and wish to compute a Cholesky factorization in finite precision arithmetic with precision $u_h > u$.
The most practically important cases are where 
$(u_h, u) = \mbox{(half, single)}$,
(half, double), or (single, double).
Naively, we might first form $A^{(h)} = fl_h(A)$,
where $fl_h$ denotes the operation of rounding to precision $u_h$,
and then compute the Cholesky factorization of $A^{(h)}$ in precision $u_h$.
For half precision this procedure can fail for two reasons. 
First, if fp16 is used then the limited range might cause overflow during
the rounding. Second, for both bfloat16 and fp16, $A^{(h)}$ might not be
(sufficiently) positive definite, because a matrix whose smallest eigenvalue is safely bounded
away from zero with respect to single precision or 
double precision may become numerically indefinite
under the perturbation induced by rounding to half precision.  
The second issue can also be encountered when a double precision matrix is
rounded to single precision. To overcome these problems we will use scaling
and shifting.

\subsubsection{Scaling}

The first step is to scale the matrix $A$ to
$H = D^{-1} A D^{-1}$, 
$D = \smash{\text{diag}(a_{ii}^{1/2})}$, and 
$D$ will be kept at precision~$u$.
Because Cholesky factorization is essentially numerically invariant under two-sided
diagonal scaling 
(as can be seen from the recurrence relations for the
Cholesky factors), the sole reason for scaling
is to reduce the dynamic range in order to avoid overflow and reduce the
chance of underflow, for fp16.  For bfloat16 or single precision it will
not usually be necessary to scale, and we can work with $A$ throughout.  
For the rest of
the presentation we will always scale, for simplicity of notation. 
Two-sided diagonal scaling before rounding to low precision was used in
\cite{hpz19} in the context of LU factorization.  
The scaling used there equilibrates rows and columns; our scaling with
$D$ is the natural analogue of that for symmetric positive definite  matrices.
For Cholesky factorization we need an extra
ingredient to ensure a successful factorization, which we consider next.

\subsubsection{Shifting}

We now convert $H$ to the lower precision $u_h$, incorporating a 
shift to ensure that the lower precision matrix is sufficiently positive definite for 
Cholesky factorization to succeed, as discussed in \cite[Sec. 2]{hipr19a}.
We will shift $H$ by $c_n u_h I$, where $c_n$ is a positive integer
constant, to obtain $G = H + c_nu_h I$.
Since the diagonal of $H$ is $I$, this shift incurs no rounding error and it produces the same result whether we shift in precision $u$ then round or round then shift in precision $u_h$.

For fp16, in view of the narrow range we will also multiply the shifted
matrix by a scalar to bring it close to the overflow level $x_{\text{max}}$, in
order to minimize the chance of underflow and of subnormal numbers being
produced.  
So our final precision-$u_h$
matrix is constructed as
\begin{align}
     G &=  H + c_n u_h I, \nonumber\\
     \beta &= 1 + c_n u_h, \quad \mu = \theta x_{\text{max}}/\beta, \\
    A^{(h)} &= fl_h( \mu G), \nonumber
\end{align}
where $\theta \in (0,1)$ is a parameter. 
Note that $\beta = \max_{ij} |g_{ij}|$,
so the largest absolute value of any element of $A^{(h)}$ is $\theta x_{\text{max}}$.
Note also that since the growth factor
for Cholesky factorization is $1$ (see, e.g., \cite[Prob.~10.4]{higham2002accuracy}), there is no
danger of overflow during Cholesky factorization of $A^{(h)}$. 

We refer to \cite[Sec. 3.3]{hipr19a} for an analysis regarding the
choice of $c_n$. However since the estimates are pessimistic, 
we take the pragmatic approach of taking $c_n$ to be a small
constant $c$.
If the Cholesky factorization fails we will increase $c$ and try again. 
We will determine experimentally how large $c$ should be for a range of 
problems of interest. Based on this we give the low precision
Cholesky factorization algorithm in Algorithm~\ref{alg.Chol-half}.

\begin{algorithm}
\caption{(Cholesky factorization in precision $u_h$). 
Given a symmetric positive definite $A\in\mathbb{R}^{n \times n}$ in precision $u$ 
this algorithm computes an approximate
Cholesky factorization $R^TR \approx \mu D^{-1} A D^{-1}$ at precision $u_h$,
where $D = \text{diag}(a_{ii}^{1/2})$.
The scalar $\theta\in(0,1]$ and the positive integer $c$ are parameters. 
}
\label{alg.Chol-half}
\begin{algorithmic}[1]
\State\label{line.chol1} 
$D = \text{diag}(a_{ii}^{1/2})$, $H = D^{-1}AD^{-1}$ 
\quad \% Set $h_{ii}\equiv 1$ instead of computing it.
\State\label{line.2}%
$G = H + c u_h I$ \label{line.c}
\State
$\beta = 1 + c u_h$
\label{line.beta}%
\State $\mu = \theta x_{\text{max}} /\beta$
\label{line.Ah}
\State $A^{(h)} = fl_h( \mu G )$ \label{line.fp16-matrix}
\State
Attempt Cholesky factorization $A^{(h)} = R^TR$ in precision $u_h$.
\If {Cholesky factorization failed}
\State $c \gets 2c$, goto line~\ref{line.2}
\EndIf
\end{algorithmic}
\end{algorithm} 

\subsection{Iterative Refinement for Least Squares Problems}


We consider the linear least squares
problem
$\min_x \Vert Ax-b\Vert_2$, where $A\in\mathbb{R}^{m \times n}$ with $m\ge n$ has full rank. 
The ideas of mixed-precision 
iterative refinement and GMRES-IR
can be adapted to the least squares case.
Least squares problems may be ill conditioned in practice, and so rounding  errors may result in an insufficiently accurate solution. In this case, 
iterative refinement may be used to improve accuracy,
and it also improves stability.  

\subsubsection{Cholesky-Based Approach}
The normal equations method solves
$$
   A^TA x = A^Tb
$$
using the Cholesky factorization of $A^TA$ (see Section \ref{sec.lp-chol}). 
In general, this method is deprecated by numerical analysts because it 
has a backward error bound of order $\kappa_2(A)u$ \cite[sect.~20.4]{higham2002accuracy} 
and the Cholesky factorization 
can break down for $\kappa_2(A) > u^{-1/2}$,
but it is used by statisticians with some justification
\cite{hist87}.
Here, we assume that $A$ is (sufficiently) well conditioned.
We propose the GMRES-IR-based least squares solver given in
Algorithm \ref{alg.ls-GMRES-IR}.

\begin{algorithm}
\caption{(Cholesky-based GMRES-IR for the least squares problem) 
Let a full rank $A\in\mathbb{R}^{m \times n}$, where $m \ge n$, 
and $b\in\mathbb{R}^m$ be given in precision $u$. This algorithm solves 
the least squares problem $\min_x\Vert b-Ax\Vert_2$ using 
Cholesky-based GMRES-IR\@. 
The scalar $\theta\in(0,1]$ and the positive integer $c$ are parameters.}
\label{alg.ls-GMRES-IR}

\begin{algorithmic}[1]
\State Compute $B = AS$, where $S = \text{diag}(1/\Vert a_j\Vert_2)$,
with $a_j$ the $j$th column of $A$.\label{line.ls1}%
\State $\mu = \theta x_{\text{max}}$
\State $B^{(h)} = fl_h(\mu^{1/2} B)$ 
\State Compute $C = B^{(h)T} B^{(h)}$ in precision $u_h$. \label{line.fp16-normal-mat} 
\label{line.ls2}%
\State Compute the Cholesky factorization $C + c u_h \text{diag}(c_{ii}) = R^TR$ 
in precision $u_h$. \label{line.cLS}
\label{line.ls3}%
\If {Cholesky factorization failed}
\State $c \gets 2c$, goto line~\ref{line.ls3}
\EndIf
\State Form $b^{(h)} = fl_h(S A^T b)$.
\State Solve $R^TR y_0 = b^{(h)}$ in precision $u_h$ and form $x_0 = \mu S y_0$
at precision $u$.
\For{$i=0 \colon i_{\max}-1$} \label{line.ir-loop}
\State Compute $r_i = A^T(b-Ax_i)$ at precision $u_r$ and round $r_i$
       to precision $u$.
\State{\label{line.normal-ir}%
  Solve $MA^TA d_i = M r_i$ by \ac{gmres} 
at precision $u$, where $M = \mu S R^{-1} R^{-T} S$ and 
matrix--vector products with $A^TA$ 
are computed at precision $u_r$, and store $d_i$ at precision $u$.}
\State $x_{i+1}=x_i+d_i$ at precision $u$.
\If {converged}
\State return $x_{i+1}$, \textbf{quit}
\EndIf
\EndFor
\end{algorithmic}
\end{algorithm} 

We make some comments on the algorithm.
Line~\ref{line.ls1} produces a matrix $B$ with columns of unit 2-norm.
The computation $C = {B}^{(h)T} B^{(h)}$ on line~\ref{line.ls2}
produces a symmetric positive definite matrix with constant diagonal elements $\mu = \theta x_{\text{max}}$,
so overflow cannot occur for $\theta < 1$.
The shift on line~\ref{line.ls3} is analogous to that in
Algorithm~\ref{alg.Chol-half},
but here the matrix $C$ is already well scaled and 
in precision $u_h$ so there is no need
to scale $C$ to have unit diagonal.

There are two reasons why explicitly forming 
$C = {B}^{(h)T} B^{(h)}$ in Algorithm~\ref{alg.ls-GMRES-IR} is reasonable
from the numerical stability point of view.
First, $C$ is used to form a preconditioner, so  the usual problems with forming a cross product matrix (loss of significance and condition squaring)
are less of a concern.
Second, if we are working in fp16 on an NVIDIA V100 we can exploit the
tensor cores when forming $C$ to accumulate block fused multiply-add
operations in single precision; this leads to a more accurate $C$, as shown
by the error analysis of Blanchard et al.\ \cite{bhlm19}.

For the computed $\hat{R}$ we have 
$$
    \hat{R}^T\hat{R} \approx B^{(h)T} B^{(h)} \approx \mu S A^TA S,
$$
or 
$$
 (A^TA)^{-1} \approx \mu S\hat{R}^{-1}\hat{R}^{-T} S,
$$
so we are preconditioning with an approximation to the inverse of $A^TA$. 
We apply the preconditioned operator $MA^TA$ to vectors at precision $u_r$.
Computing $y = A^TA x$ costs $4mn$ flops 
and $SR^{-1}R^{-T}y$ costs another $2n^2+n$ flops, 
making $4mn + 2n^2 + n$ flops in total.
For $m \gg n$ and large $n$, computing $y = A^TA x$ costs a factor $n/4$
fewer flops than the $mn^2$ flops needed to form $A^TA$, 
while for $m\approx n$ the difference is a factor $n/6$.
For large $n$, 
even allowing for the fact that the flops we are comparing are at different precisions,
as long as \ac{gmres} converges quickly the cost of 
the refinement stage should be negligible compared with the cost of 
forming $A^TA$ and computing the Cholesky factorization.

Related to this work is the Cholesky--QR algorithm for computing a QR factorization $A = QR$.
It forms the cross-product matrix $A^TA$, computes the Cholesky factorization $A^TA = R^TR$,
then obtains the orthogonal factor $Q$ as $Q = AR^{-1}$,
and this process can be iterated for better numerical stability;
see, for example, 
\cite{fkny18}, 
\cite{ynyf15}, 
\cite{ytd15}, \cite{ytd16}.
Our work differs in that we do not compute $Q$, we carry out the Cholesky factorization 
in lower precision than the working precision,
and we solve a least squares problem using preconditioned iterative refinement.

\subsubsection{Augmented Matrix Approach}
Another approach to mixed precision least squares iterative refinement 
was presented by Carson, Higham, and Pranesh in~\cite{chp20}. This approach is 
based on the method of 
using the QR factorization 
\begin{equation*}
A=Q
\begin{bmatrix}
R \\ 0
\end{bmatrix},
\end{equation*}
where $Q=[Q_1,Q_2]\in\mathbb{R}^{m\times m}$ is an orthogonal matrix with 
$Q_1\in\mathbb{R}^{m\times n}$ and $Q_2\in\mathbb{R}^{m\times (m-n)}$, and 
$R\in \mathbb{R}^{n\times n}$ is upper triangular.
The unique least squares 
solution is $x=R^{-1}Q_1^T b$ with residual $\Vert b-Ax\Vert_2 = \Vert Q_2^Tb\Vert_2$. 

An iterative refinement approach that works even when $Ax=b$ is
inconsistent was suggested by Bj{\"o}rck~\cite{bjorck67}. Refinement is
performed on the augmented system
\begin{equation}
\begin{bmatrix}
I & A \\
A^T & 0
\end{bmatrix}
\begin{bmatrix}
r \\
x
\end{bmatrix}
=
\begin{bmatrix}
b \\
0
\end{bmatrix},
\label{augsys}
\end{equation}
which is equivalent to the normal equations. 
In this way, the solution $x_i$ and residual $r_i$ for the least squares problem 
are simultaneously refined. 
Bj{\"o}rck~\cite{bjorck67} shows that the linear
system can be solved by reusing the QR factors of $A$. 

Existing analyses of the convergence and accuracy of this approach in 
finite precision assume that at most two precisions are used; the 
working precision $u$ is used to compute the QR factorization, 
solve the augmented system, and compute the update. A second precision $u_r \leq u$ is used to 
compute the residuals. Typically $u_r = u^2$, in 
which case it can be shown that as long as the condition number of 
the augmented system matrix is smaller than $u^{-1}$, the refinement 
process will converge with a limiting forward error on the order 
of $u$; see~\cite{bjorck90} and~\cite[sect.~20.5]{higham2002accuracy} and 
the references therein.   

The work~\cite{chp20} shows that the three-precision iterative refinement 
approach of Carson and Higham~\cite{carson2018accelerating} can be applied in this case; the theorems developed 
in~\cite{carson2018accelerating} regarding the forward error and 
normwise and componentwise backward error for iterative refinement of
linear systems are applicable. The only thing that must change is the
analysis of the method for solving the correction equation since we now work with a QR factorization of $A$, which  
can be used in various ways. 

The work in~\cite{chp20} also extends the \ac{gmres}-based refinement scheme of~\cite{carson2017new} 
to the least squares case and shows that one can construct a left preconditioner 
using the existing QR factors of $A$ such that \ac{gmres} provably converges to a 
backward stable solution of the preconditioned augmented system. Further, 
it is shown that an existing preconditioner 
developed 
for saddle point systems can also work well in the \ac{gmres}-based approach in practice, 
even though the error analysis is not applicable. We refer the reader to~\cite{chp20} for 
further details.

\subsection{Quantized Integer LU Factorization}

Quantization is a technique widely being used in deep learning
inference~\cite{pytorch1.4quantization, tensorflow8bitquantization}.
While the model is usually still trained in single precision,
quantization compress the data and use lower precision to carry
out the computation in inference stage which is applying the
trained model to new data for real application.
For an int8 quantized model, the data is converted into
8-bit integers. The computation and communication are reduced
4 times comparing to 32-bit single precision while the accuracy lost
is acceptable (usually $<1\%$ for predictive models).
Integer arithmetic is available on most hardware architectures.
\ac{fpga}s are usually more capable in integer operations and might not
have floating-point number arithmetic units. New \ac{asic}s for deep
learning inference are also moving toward using mostly integer
arithmetic for quantized neural networks. This motivated to investigate
the use of integer arithmetic for the Gaussian
elimination (LU factorization) with partial pivoting.

\subsubsection{Quantized Integer LU Algorithm}

\begin{table}[h!]
\centering
\begin{tabular}{|Sc|Sc|}
 \hline
Storage format & $i$ in 32-bit integer\\ \hline
Represented real number & $R(i)=i/2^{32}\times 2^0$ \\ \hline
Conversion from \texttt{double} precision number $\alpha$
& $i \leftarrow \texttt{int32}(\alpha \times 2^{32})$ \\ \hline
Conversion to \texttt{double} precision number $\alpha$
& $\alpha \leftarrow \texttt{double}(i) / 2^{32}$ \\ \hline
Addition  & $R(i)+R(j)=i/2^{32}+j/2^{32}=(i+j)/2^{32}=R(i+j)$ \\ \hline
\multirow{2}{*}{Multiplication} & $R(i)\times R(j)=i/2^{32}\times j/2^{32}=(i\times j)/2^{64}$ \\
& $=(i\times j/2^{32})/2^{32}=R(i\times j / 2^{32})$ \\
 \hline
\end{tabular}
\caption{Proposed Fixed-point Number Representation}
\label{table:qilu_representation}
\end{table}

The basic idea is to scale down numbers to fit into a fixed-point number
representation: $i/2^{32}\times 2^0$ where $i$ is in 32 bits integer.
The exponent will not change under addition or multiplication so can be
ignored. The addition under is form is simply integer addition.
Multiplication becomes: $i/2^{32} \times j/2^{32} = i\times j
/2^{64}=(i\times j/2^{32})/2^{32}$. To compute $i\times j/2^{32}$ can be
done with 32 bits integer multiply and return the high 32 bits in the 64
bits result.  Note that this operation can be done in one instruction on
modern CPU instruction set architectures (ISAs) including \texttt{x64}
and \texttt{ARM}. Table~\ref{table:qilu_representation} summarizes the
proposed fixed-point number representation.

\begin{algorithm}[tb]
\begin{algorithmic}[1]
\State
\begin{tabular}{l l}
\textbf{Input}: &$n$ by $n$ matrix $A$ in double
precision. \\
& Integer $r$ for the range while normalizing $A$.
\end{tabular}
\State Declare identity matrix $P$ as permutation matrix.
\State $m \leftarrow \max(A)\times2^{r}$; $A \leftarrow A / m$ \Comment{Normalize $A$ into $[-2^{-r}, 2^{-r}]$}
\State $A_{int} \leftarrow \text{int32}(A\times 2^{32})$ \Comment{Convert $A$ into proposed fixed-point representation.}
\For{$i = 1 \ldots n$} \Comment{Main loop over columns}
\State $\text{pivot} \leftarrow \left( \arg\max|A_{int}[i{:}n,i]| \right) +i-1$ \Comment{Find the pivot index.}
\State $\text{swap} \left( A_{int}[i,:], A_{int}[\text{pivot},:] \right) $ \Comment{Swap rows.}
\State $\text{swap} \left( P[i,:], P[\text{pivot},:] \right) $
\State $\alpha \leftarrow \text{int64}(2^{32}) / A[i,i]$ \Comment{Find the scale with integer division.} \label{lst:qilu:line:div}
\State $A_{int}[i{:}n, i] \leftarrow \alpha A_{int}[i{:}n, i]$ \Comment{Scale the column.} \label{lst:qilu:line:scale}
\State $A_{int}[i+1{:}n, i+1{:}n] \leftarrow A_{int}[i+1{:}n, i+1{:}n] - A_{int}[i+1{:}n, i] \times A_{int}[i, i+1{:}n] / 2^{32}$ \label{lst:qilu:line:update}
\State
\Comment{Integer rank-1 update with a division using integer shift.}
\EndFor
\State $L \leftarrow$ lower triangular part of $\text{double}(A)/2^{32}$ with unit diagonal.
\State $U \leftarrow$ upper triangular part of $\text{double}(A)/2^{32}$ including diagonal.
\State \textbf{Return}: $P, L, U$ as the result of factorization such that $P (A/m)=L U$
\end{algorithmic}
\caption{LU factorization with partial pivoting based on integer arithmetic.}
\label{alg:qilu}
\end{algorithm}

Algorithm \ref{alg:qilu} shows for LU factorization with partial pivoting
based on integer arithmetic. The computation inside the loop is mainly
32-bit integer arithmetic. Line~\ref{lst:qilu:line:div} requires
64-bit integer division but only once per column.
The scale in line~\ref{lst:qilu:line:scale} will remain in
\texttt{int32} range because the pivot has larger magnitude then other
elements in the column. The update in line~\ref{lst:qilu:line:update}
is 32-bit integer multiply but we only need the high 32 bits
in 64 bits results.

The input integer $r$ determines the number of bits ($32-r$) we are actually using while converting $A$ into integer.
Because the matrix would grow during the factorization and
we do not have any dynamic scaling during the factorization,
it might hit the integer range and overflow at some point.
To avoid it, we first scale the matrix into $[-2^{-r}, 2^{-r}]$.
The higher $r$ is, the more room we will have from the integer range.
But less accurate the input matrix would be after converted into
\texttt{int32}.

\subsubsection{Quantized Integer LU Numerical Results}

\begin{figure}[tb]
\centering
\includegraphics[width=0.8\linewidth]{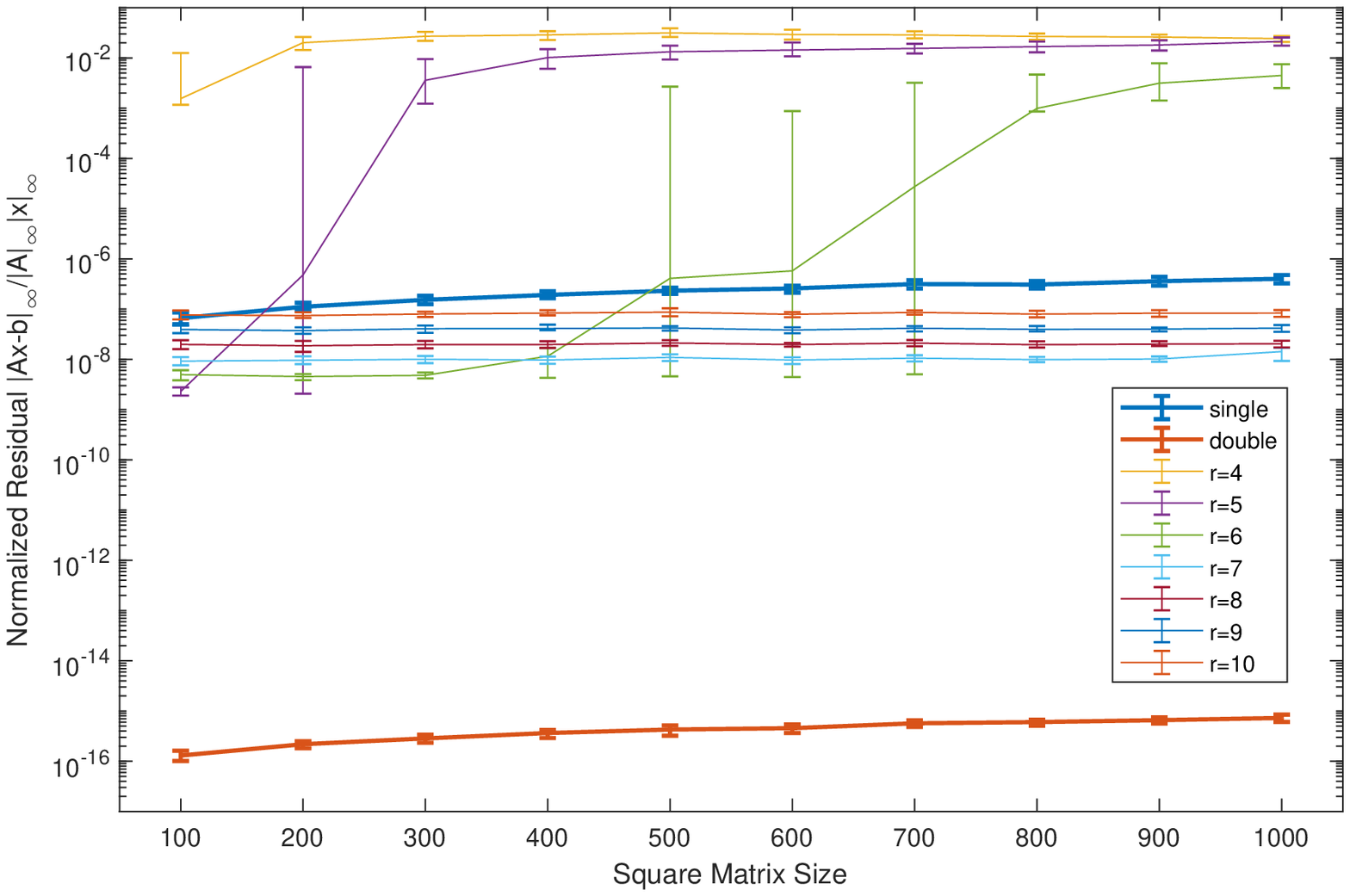}
\caption{Normalized backward error vs. input matrix size for different number of usable bits $r$. Result in single and double precision are also shown.}
\label{fig:qilu_bckerr}
\end{figure}

Figure \ref{fig:qilu_bckerr} shows the normalized backward error
$\| Ax-b \|_\infty /\|A\|_\infty \|x\|_\infty$  vs. input matrix size.
The algorithm is implemented in \texttt{MATLAB R2018b}. Each element
of the matrix is generated from uniform random distribution:
\texttt{uniform}(-1,1). Each point is the the geometric average over
30 random matrices and error bars indicate the 15\% and 85\% percentiles.
The result from single and double precision LU factorization is
also reported in bold lines as reference.
For the results which are $>10^{-6}$, overflow occurred and the algorithm failed.
Otherwise there's no numerical error during the factorization.
The error is only introduced in the conversions between floating-point
and fixed-point format, not during integer factorization.
The backward error grows with $r$ since the input is
truncated more at the conversion. But still when $r=10$ it is still using
$32-10=22$ bits and the result is comparable with single
precision which is using 23 mantissa bits.

\subsubsection{Future Work on Quantized Integer LU}

We would like to show case that it is possible to have low precision
factorization using integer arithmetic. For the next step we will
conduct more detailed error analysis. Extend to shorter integers
such as \texttt{int16} and \texttt{int8}. Also to tackle the overflow 
problem, we would like to consider dynamically scale the columns during
factorization to keep the numbers in range. And the same as per channel
quantization in deep learning, assign a different range to each column
might also be a feasible approach.

\subsection{Symmetric Eigenvalue Problems}

In \cite{DMW}  an algorithm is described for determining rigorous
error bounds for a simple eigenvalue and its associated eigenvector.
The algorithm has the pleasing feature of providing an improved
eigenpair as a by-product. This approach assumes that an
eigenpair is given. No assumptions are made about how that eigenpair
was found, whether through some knowledge of the physical problem, an
initial eigenvalue decomposition in a lower precision or a clever guess.

We are interested in improving the accuracy of an eigenvalue-
eigenvector pair. Consider the eigenvalue problem $Ax = \lambda x$, where $\lambda$ and
$x$ have been found by some means. Because they were arrived at by some
calculation on a computer with finite precision or by some insight into
the problem, they are, in general, not the true eigenvalue and eigenvector,
but an approximation. We know, however, that there exist $\mu$
and $\tilde {y} $ such that
$A(x+\tilde{y} =\lambda + \mu) = (\lambda + \mu ) (x + \tilde{y}) $
is the exact solution to the eigenvalue problem, where $\mu$ and $\tilde{y}$ are the
corrections to the computed $\lambda$ and $x$.

We will normalize $x$ such that $||x|| _ \infty $= 1 and say $x _s  = 1$, where the $s ^{th}$
component of $x$ is the largest. This can be done because we have one
degree of freedom in our choice of the components for $x$. We will assume
that the $s ^{th}$ component of $x$ is exact and no correction is needed. This
determines the value of $\tilde{y} _s $, which is the correction to $x_s$. Because $x_s$
is exact, the value of $\tilde{y} _s $ is zero. This also determines the degree of
freedom in the corrected vector, $x+\tilde{y}$, through the relationship between $x$
and $\tilde{y}$, namely $(x+\tilde{y}_s )= 1$.

We can rewrite equation
$A(x+\tilde{y} =\lambda + \mu) = (\lambda + \mu ) (x + \tilde{y}) $ as
$(A-\lambda I)\tilde{y} - \mu x = \lambda x-Ax + \mu \tilde{y}$.
Note that $\lambda x-Ax$ is the residual for the computed eigenvalue and eigenvector.
If we look more closely at the product $(A-\lambda I)\tilde{y} $, we discover
that because $\tilde{y}_s = 0$, the $s ^{th}$ column of $(A-\lambda I)$ does not participate in the
product with $\tilde{y}$. In the formulation of $(A-\lambda I)\tilde{y}- \mu x$, we can replace the $s$
component of $\tilde{y}$, which is zero, by the value $\mu$ and the $s ^{th}$ column of $(A-\lambda I)$
by $-x$ to arrive at $(A-\lambda I)\tilde{y} - \mu x $.

We will define $y$ by $y \equiv \tilde{y} + \mu e_s$, where $e_s$ is the $s ^{th}$ column of the
identity matrix. So the $s ^{th}$ component of the newly defined $y$ has the
value $\mu$; i.e., $y_s = \mu$. We will also define the matrix $B$ as the matrix
$(A-\lambda I)$ with the $s ^{th}$ column replaced by $-x$. Thus we can rewrite
$(A-\lambda I)\tilde{y} - \mu x = \lambda x-Ax + \mu \tilde{y}$
as
$By = r + y_s \tilde{y}$,
where $r = \lambda x-Ax$.

Because the $n+1$ element of the solution vector is known, we
will solve with the truncated form of $B$, truncated so the n+l row and
n+l column are no longer present. This truncation can be done because
we know the solution vector has a zero in the $(n+1)^{th}$ position.
The above equation is a nonlinear equation defining the correction $y$.
This system can be solved by the following iterative method for solving,

$By^{(p+1)} = r+ y^{(p)}_s \tilde{y}^{(p)}$, where
$\tilde{y}^{p} = y_s ^{(p)} -  y_s ^{(p)} e_s$.
This is the approach used in \cite{Dongarra:1982:sicedr}.

\begin{algorithm}
\begin{algorithmic}[1]
\State
  \textbf{Input}: $A = A^T \in \mathbb{R}^{n \times n}$, $\widehat{X} \in
  \mathbb{R}^{n \times \ell}$, $1 \le \ell \le n$
\State
  \textbf{Output}: $X' \in \mathbb{R}^{n \times \ell}$,
  $\widetilde{D}=\text{diag}\left( \widetilde{\lambda_i} \right) \in
  \mathbb{R}^{\ell \times \ell}$, $\widetilde{E} \in \mathbb{R}^{\ell \times
  \ell}$, $\omega \in \mathbb{R}$
\Function{$\left[ X', \widetilde{D}, \widetilde{E}, \omega \right]$ $\leftarrow$
  \textsf{RefSyEv}}{$A$, $\widehat{X}$}
  \State $R \leftarrow \mathbb{I}_n - \widehat{X}^T \widehat{X}$
  \label{alg:symeigref:line:r}
  \State $S \leftarrow \widehat{X}^T A \widehat{X}$
  \label{alg:symeigref:line:s}
  \State $\widehat{\lambda_i} \leftarrow s_{ii} / (1-r_{ii})$ $\quad$
  for $i=1,\ldots,\ell$ \Comment{Compute approximate eigenvalues.}

  \State $\widetilde{D} \leftarrow \text{diag} \left( \widetilde{\lambda_i} \right)$
  \State $\omega \leftarrow 2 \left( \left\| S - \widetilde{D} \right\|_2 + \|A\|_2
  \|R\|_2 \right)$ \label{alg:symeigref:line:norm}
  \State $e_{ij} \leftarrow \begin{cases} \frac{s_{ij}+\widetilde{\lambda}_j
  r_{ij}}{\widetilde{\lambda}_j-\widetilde{\lambda}_i}  & \text{if }
  \left| \widetilde{\lambda}_i-\widetilde{\lambda}_j \right| > \omega \\
  r_{ij}/2 & \text{otherwise} \end{cases}$ for $1 \le i,j \le \ell$
  \Comment{Compute the entries of the refinement matrix $\widetilde{E}$.}
  \label{alg:symeigref:line:e}
  \State $X' \leftarrow \widehat{X} + \widehat{X} \widetilde{E}$
  \label{alg:symeigref:line:x}
  \Comment{Update $\widehat{X}$ by $\widehat{X}(\mathbb{I}_n+\widetilde{E})$}
\EndFunction
\end{algorithmic}
\caption{Iterative refinement for symmetric eigenvalue problem.}
\label{alg:symeigref}
\end{algorithm}

\begin{figure}[tb]
\centering
\includegraphics[width=0.5\linewidth]{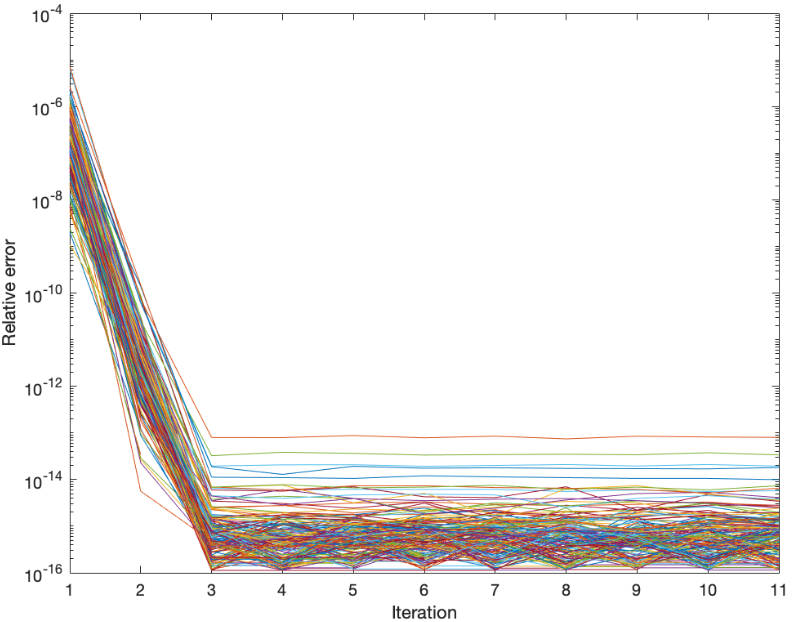}
\caption{Convergence of eigenvalue refinement from single to double
  precision for $n=200$.}
\label{fig:refsyev_conv}
\end{figure}

Algorithm~\ref{alg:symeigref} shows another approach using an iterative refinement procedure
for solving a symmetric eigenvalue problem~\cite{Ogita:2018:itrrefeig}. This methods succeeds also for clustered eigenvalues~\cite{Ogita:2019:clstrdeigs}.
Line~\ref{alg:symeigref:line:r},~\ref{alg:symeigref:line:s}, and~\ref{alg:symeigref:line:x} represent the compute intensive parts of
the algorithm, which amounts to 4 calls to matrix-matrix multiply
function \texttt{xGEMM}. Line~\ref{alg:symeigref:line:norm}
is the matrix norm.
The original analysis uses 2-norm but it is suggested to approximate
it using the Frobenius norm because it is much easier to
compute in practice. Line~\ref{alg:symeigref:line:e}
is an element-wise operation to construct
the refinement matrix $E$. Line~\ref{alg:symeigref:line:x}
is the update of eigenvectors by
applying the refinement matrix $E$. High-precision
arithmetic is required for all computations except
line~\ref{alg:symeigref:line:norm}
for matrix norm. Although the algorithm can work
for only a subset of eigenvectors, it is only refining in the corresponding
subspace. Hence the refinement process could be limited.
In other words, the desired accuracy might be unattainable if only a part of
the spectrum is refined in higher precision.

Figure~\ref{fig:refsyev_conv} shows the convergence behavior of
algorithm~\ref{alg:symeigref} on a real symmetric matrix of size $n=200$ when
refining the entire eigen-specturm.
Each line represents the convergence of one eigenvalue,
and the normalized residual $\|Ax-\lambda x\|/\|A\| \|x\|$ is
plotted against subsequent iteration numbers.

Iterative refinement based on linear solve is also
possible~\cite{Dongarra:1982:sicedr}. Algorithm~\ref{alg:sice} is
the procedure called SICE which, in each iteration, solves a linear system
resulting from a rank-1 update in order to refine a single
eigen-pair.  The rank-1 update is introduced while replacing one column
in $A-\lambda I$ to remove one degree of freedom on eigenvector
correction and, at the same time, compute a correction for the
corresponding eigenvalue.  The original
formulation~\cite{Dongarra:1982:sicedr} solves the system with two
series of Givens rotations to make it upper triangular. This process is
hard to parallelize on modern architectures. Also, some form of
orthogonalization should be considered while using the algorithm to
refine more than one eigenvalue.

\begin{algorithm}
\begin{algorithmic}[1]
\Function{$[x, \lambda]$ $\leftarrow$
\textsf{SICE}}{$A$, $x_0$, $\lambda_0$}
\State $[Q, T] \leftarrow$schur$(A)$ \Comment Schur decomposition
to find $A=QTQ^T$ where T is quasi upper triangular.
\State $[m, s] \leftarrow max(x_0)$
\Comment Find maximum value and index in the eigenvector. 
\State $x_0 \leftarrow x_0/m$ \Comment Normalize
\For{ $i = 1, 2, \ldots$ }
\State $c \leftarrow -x_{i-1}-(A-\lambda_{i-1} I)[:,s]$
\Comment Column $s$ of $A-\lambda_{i-1} I$
\State $d \leftarrow Q^Tc$
\State $f \leftarrow e_s^T Q$ \Comment Row $s$ of $Q$
\State Solve the rank-1 updated system $ Q(T-\lambda_{i-1} I + df^T) Q^T y_i = Ax_{i-1}-\lambda_{i-1} x_{i-1} $
\State $\lambda_i \leftarrow \lambda_{i-1} + y_i[s]$
\Comment Eigenvalue correction.
\State $x_i \leftarrow x_{i-1} + y_i$ \Comment Eigenvector correction.
\State $x_i[s] \leftarrow x_{i-1}[s]$ \Comment Restore x[s].
\If{ $2\times y_i[s] > y_{i-1}[s]$ }
\State Break from for loop.
\EndIf
\EndFor
\State $x \leftarrow x_i$
\State $\lambda \leftarrow \lambda _i$
\EndFunction
\end{algorithmic}
\caption{SICE algorithm for iteratively refining computed eigenvalue.}
\label{alg:sice}
\end{algorithm}

In many applications, we are satisfied with a subset of the eigenvalue
eigenvector pairs. In this case, it can be much more efficient to use an
algorithm such as the Multiple Relatively Robust Representations (MRRR)
\cite{Dhillon:2006:mr3} to compute the eigenpairs once the matrix has been
reduced to a tri-diagonal form. Though this method by itself is less
accurate than its counterparts (Divide and Conquer and QR),
\cite{2014:Petschow:mr3} show that using a mixed precision approach can be
beneficial to improve the accuracy of the solve and the overall time to
solution. The mixed precision approach here also shows promise in improving the
orthogonality over its single precision and other solver counterparts.

%% file: lowp_blas.tex
\subsection{Low Precision BLAS}
The revolution of machine learning applications and artificial intelligence (AI) spiked an interest in developing high-performance half-precision arithmetic ($16$-bit floating-point format), since most AI applications do not necessarily require the accuracy of single or double precision~\cite{gupta2015deep}. Half precision also enables machine learning applications to run faster, not only because of the faster arithmetic, but also because of the reduction 
in memory storage and traffic by a factor of $2\times$ against single precision, and by a factor of $4\times$ against 
double precision. 

In terms of vendor support, NVIDIA, Google, and AMD manufacture hardware that is capable of performing floating point arithmetic using $16$-bit formats. Google's Tensor Processing Units (TPUs) are customized chips that are mainly designed for machine learning workloads using the \texttt{bfloat16} format. AMD also provides half-precision capabilities, and their software stack shows support for both the \texttt{bfloat16} format and the IEEE format~\cite{ieee754}. The theoretical performance of half-precision on AMD GPUs follows the natural $2\times$/$4\times$ speedups against single/double precisions, respectively. As an example, the Mi50 GPU has a theoretical FP16 performance of $26.5$ Tflop/s, against a $13.3$ Tflop/s for FP32 and $6.6$ Tflop/s for FP64. But perhaps the most widely accessible hardware with half-precision capability are NVIDIA GPUs, which have introduced half-precision arithmetic since the Pascal architecture. Throughout this section, we will focus on NVIDIA GPUs and its math libraries to highlight half-precision developemnts for numerical kernels. 

NVIDIA GPUs implement the ``binary16'' format which is defined by the IEEE-754 standard~\cite{ieee754}. While the Pascal GPU architecture introduced hardware support for FP16 arithmetic, the Volta architecture, which powers the 
Summit supercomputer,\footnote{https://www.olcf.ornl.gov/summit/} comes with hardware acceleration units (called Tensor Cores) for matrix multiplication in FP16. These Tensor Cores are theoretically $12\times$ faster than the theoretical FP16 peak performance of the preceding architecture. Applications taking advantage of the Tensor Cores can run up to $4\times$ faster than using the regular FP16 arithmetic on the same GPU. The Tensor Cores are also able to perform a mixed-precision multiplication, with a low precision input (e.g. half-precision) and a higher precision output (typically single-precision). The Tensor Core units are discussed in more details in Section~\ref{subsubsec:tensor_cores}.

In terms of half-precision Basic Linear Algebra Subroutines (BLAS), most of the available 
routines consider only dense matrix multiplications (GEMMs). From the perspective of machine learning applications, most of the performance critical components in training/inference can be reformulated to take advantage of the GEMM kernel. As for dense linear algebra, many high level algorithms are built to extract their high performance from GEMM calls. Therefore, accelerating such performance-critical steps through FP16 GEMM (HGEMM) would propagate the performance advantage to the entire algorithm, while keeping other numerical stages in their original precision(s). An example of this practice is the mixed precision dense LU factorization~\cite{haidar2018harnessing}, which is used to accelerate the solution of $Ax=b$ in double precision, see~\Cref{sec:denseiterref}. 

\subsubsection{Hardware Acceleration of Half Precision}
\label{subsubsec:tensor_cores}
The CUDA Toolkit is one of the first programming models to provide half-precision (i.e., FP16) arithmetic. Early support was added in late 2015 for selected embedded GPU models that are based on the Maxwell architecture. The FP16 arithmetic has become mainstream in CUDA-enabled GPUs since the Pascal architecture. In general, half precision has a dynamic range that is significantly smaller than single or double precision.

The Volta and Turing architectures introduce hardware acceleration for matrix multiplication in FP16. The hardware acceleration units are called Tensor Cores. They can deliver a theoretical peak performance that is up to $8\times$ faster than the peak FP32 performance. As an example, each Volta V100 GPU has $640$ Tensor Cores, evenly distributed across $80$ multiprocessors. Each Tensor Core possesses a mixed-precision $4\times 4\times 4$ matrix processing array which performs the operation $D = A\times B + C$, where $A$, $B$, $C$ and $D$ are $4\times 4$ matrices. The inputs $A$ and $B$ must be represented in FP16 format, while $C$ and $D$ can be represented in FP16 or in FP32 formats. It is also possible that $C$ and $D$ point to the same matrix. 

NVIDIA's vendor library cuBLAS provides various optimized routines, mostly GEMMs, that can take advantage of the Tensor Core acceleration if configured accordingly. As an example, the routine {\texttt{cublasHgemm}} implements the GEMM operation for real FP16 arithmetic. 

\begin{figure}[!htb]
\centering
\includegraphics[width=0.7\linewidth]{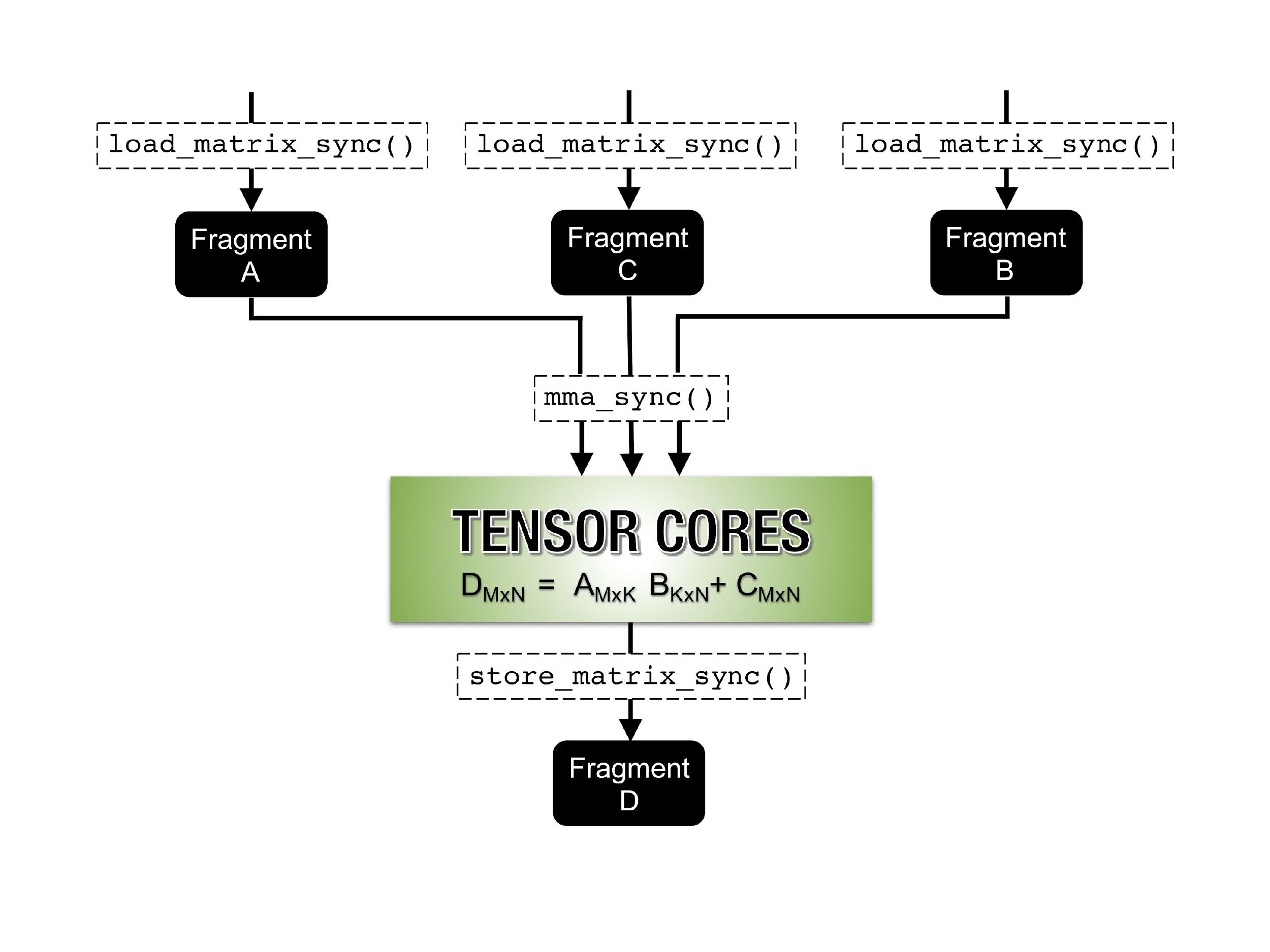}
\caption{Programmability of the Tensor Core units}
\label{fig:tc}
\end{figure}

Apart from the vendor library, taking advantage of the Tensor Cores in a custom kernel is possible also through the use of low-level APIs that are provided by the programming model. As shown in Figure~\ref{fig:tc}, Tensor Cores deal with input and output data through opaque data structures called \emph{fragments}. Each fragment is used to store one matrix. Fragments can be loaded from shared memory or from global memory using the {\texttt{load\_matrix\_sync()}} API. A similar API is available for storing the contents of an output fragment into the shared/global memory of the GPU. The {\texttt{mma\_sync()}} API is used to perform the multiplication. The user is responsible for declaring the fragments as required, and calling the APIs in the correct sequence. 

The programming model imposes some restrictions to the programming of the Tensor Cores. First, the GEMM dimensions  
($M$, $N$, $K$), which also control the size of the fragments, are limited to three discrete combinations, namely ($16$, $16$, $16$), ($32$, $8$, $16$), and ($8$, $32$, $16$). Second, the operations of load, store, and multiply must be performed by one full warp (32 threads). Finally, the load/store APIs require that the leading dimension of the corresponding matrix be multiple of $16$-bytes. As an example, a standard GEMM operation of size 
($16$, $16$, $16$) requires three {\texttt{load\_matrix\_sync()}} calls (for $A$, $B$, and $C$), one {\texttt{mma\_sync()}} call, 
and then a final {\texttt{store\_matrix\_sync()}} call to write the result. The latest CUDA version to date (10.1) provides direct access to the Tensor Cores through an instruction called {\texttt{mma.sync}}. The instruction allows one warp to perform four independent GEMM operations of size ($8$, $8$, $4$). However, using the explicit instruction may lead to long-term compatibility issues for open-source libraries as new architectures are released. 

\subsubsection{Half-precision GEMM (HGEMM)}
\label{subsubsec:hgemm}
\begin{figure}[!htb]
\centering
\includegraphics[width=0.7\linewidth]{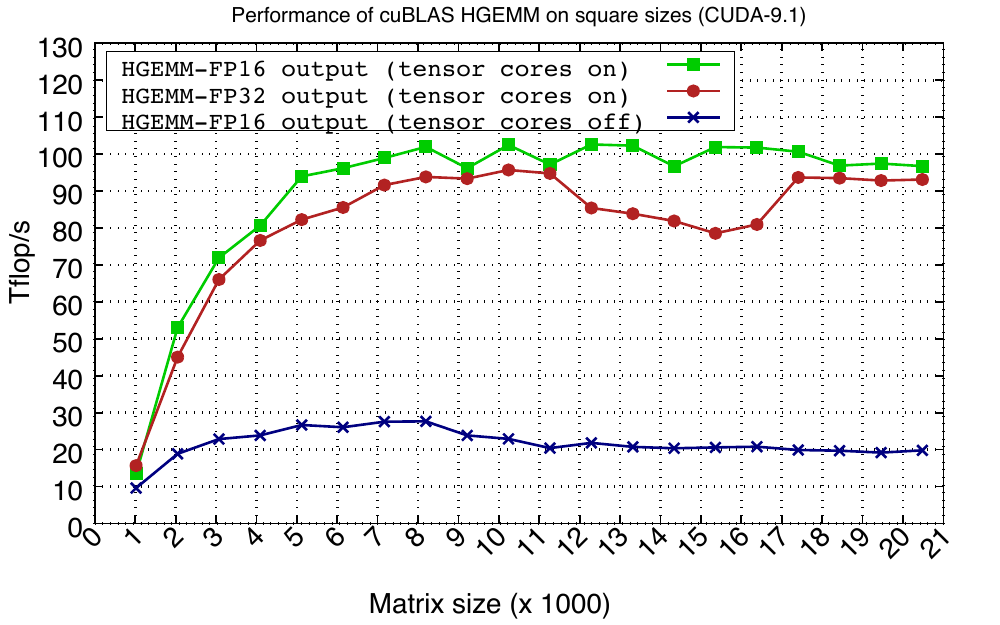}
\caption{Performance of different HGEMM kernel from the cuBLAS library on square sizes. Results are shown on a Tesla V100 GPU using CUDA-9.1.}
\label{fig:hgemm_performance}
\end{figure}

The cuBLAS library provides several routines that take advantage of the reduced
FP16 precision. Figure~\ref{fig:hgemm_performance} shows the performance of
three different HGEMM kernels. An HGEMM kernel with half-precision output can
achieve up to $30$ Tflop/s of performance if the tensor cores are turned off.
While this is around $2\times$ the single-precision performance, a significantly
higher performance can be achieved if the tensor cores are turned on. As the
figure shows, the tensor cores are capable of delivering an asymptotic $100$
Tflop/s, which is $5\times$ the asymptotic performance of a non-accelerated
HGEMM.

\begin{figure}[!htb]
\centering
\includegraphics[width=0.7\linewidth]{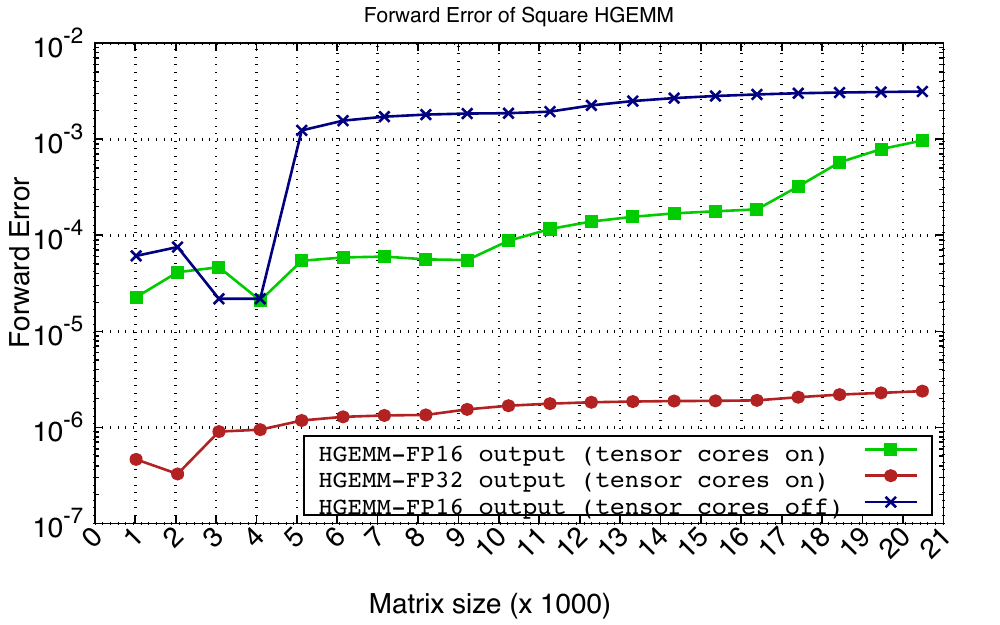}
\caption{Forward error of HGEMM with respect to MKL SGEMM ($C = \alpha AB + \beta C$). Results are shown for square sizes using cuBLAS 9.1 and MKL 2018.1. The forward error is computed as $\frac{\left \| R_{cuBLAS} - R_{MKL} \right \|_{F}}{\sqrt{k+2} \left | \alpha \right | \left \| A \right \|_F \left \| B \right \|_F + 2 \left | \beta \right | \left \| C \right \|_F}$, where $k$ is equal to the matrix size.}
\label{fig:hgemm_error}
\end{figure}

However, perhaps the most interesting performance graph of Figure~\ref{fig:hgemm_performance} is the HGEMM with FP32 output. The reason is that its performance is close to the accelerated HGEMM kernel, but with much more precision on the output. This is of particular importance for mixed-precision algorithms~\cite{haidar2018harnessing,haidar2017investigating}. To put this fact in more perspective, Figure~\ref{fig:hgemm_error} shows the forward error between the three different HGEMM kernels, with respect to the single-precision GEMM kernel from the Intel MKL library. The forward error is computed as $\frac{\left \| R_{cuBLAS} - R_{MKL} \right \|_{F}}{\sqrt{k+2} \left | \alpha \right | \left \| A \right \|_F \left \| B \right \|_F + 2 \left | \beta \right | \left \| C \right \|_F}$, where $k$ is equal to the matrix size. The first surprising observation is that an HGEMM operation with FP16 output is more accurate if the tensor cores are turned on, which means that the utilization of the tensor core units achieves both better performance and higher accuracy. The second observation is that performing HGEMM with FP32 output achieves at least two more digits of accuracy when compared with the other two HGEMM variants. Given that HGEMM with FP32 output is mostly within $90$\% of the peak tensor core throughput, it is clearly the best option for mixed-precision algorithms that target achieving higher accuracy while taking advantage of the half-precision.

\subsubsection{Batch HGEMM}
\label{subsubsec:hgemm_batched}
\begin{figure}[!htb]
\centering
\includegraphics[width=0.7\linewidth]{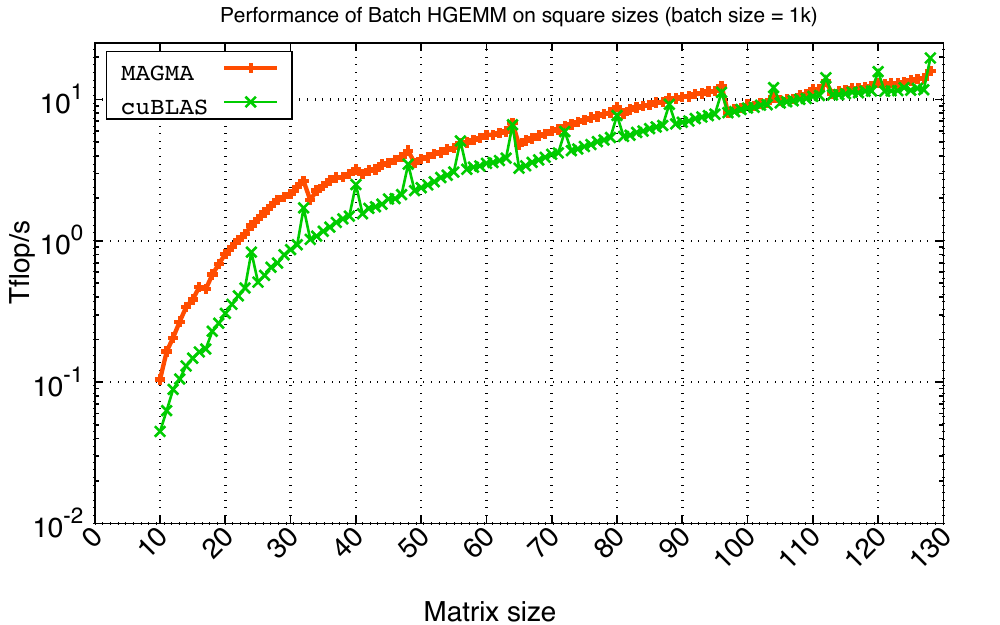}
\caption{Performance of the batch HGEMM kernel on square sizes. Results are shown on a Tesla V100 GPU using CUDA-9.1.}
\label{fig:hgemm_batched}
\end{figure}

Apart from the vendor-supplied BLAS, few efforts have focused on building open-source BLAS routines that utilize the tensor cores of NVIDIA GPUs. An example of such efforts is in the MAGMA library~\cite{gullo2009numerical}, which has a batch HGEMM kernel that makes use of the tensor cores~\cite{abdelfattah2019fast}. The kernel builds an abstraction layer over the tensor cores to overcome their size restrictions, so that arbitrary blocking sizes can be used by the kernel. The batch HGEMM kernel in MAGMA outperforms cuBLAS for relatively small sizes, as shown in Figure~\ref{fig:hgemm_batched}. The same work also shows that extremely small matrix (e.g. whose sizes $\leq 10$) may not necessarily benefit from tensor core acceleration. 

%% file: compression.tex
\section{Data and communication compression for multiprecision algorithms}
\label{sec:compression}

A fundamental requirement for accelerating scientific computations
through multiprecision use is the ability to efficiently convert data
between the different floating point formats employed and to 
minimize the communications associated with	these data movements.
Techniques and implementations to accomplish this efficiently 
have been developed usually ad-hoc, e.g., implemented and tuned 
for particular algorithms and implementations that use 
mixed-precision. Thus, although there are some solutions that address
particular challenges, there are no standards, often there are no
user-level interfaces to lower-level building blocks, and therefore
not extracted as independent, supported libraries or components
infrastructure that other developers can use. To address this, 
we have been investigating a number of building blocks that can 
be extracted and included in numerical libraries for the development 
of mixed-precision algorithms. The components that are of interest 
for data and communication compression are discussed in the 
subsequent subsections.

\subsection{Data conversions}\label{subs:casting}
Many mixed-precision algorithms need to convert data between different
standard IEEE formats. For example, LAPACK supports
this type of data conversion as needed for its mixed-precision
iterative refinement solvers. Support is provided through auxiliary
routines for either general or triangular matrices, following standard
LAPACK naming conventions and matrix representations. For example,
general matrices can be converted from FP64 to FP32 as follows:
\begin{lstlisting}[language=C++]
     zlag2c(M, N, zA, LDA, cA, LDCA, INFO)
     dlag2s(M, N, dA, LDA, sA, LDSA, INFO).
\end{lstlisting}
The first example is for casting double complex to single
complex matrix, and the second for double to single real matrix. 
The other way around (from single to double) is also provided through
the {\tt clag2z} and {\tt slag2d} routines.

The interfaces for converting triangular matrices are:
\begin{lstlisting}[language=C++]
     zlat2c(UPLO, N, zA, LDA, cA, LDCA, INFO)
     dlat2s(UPLO, N, dA, LDA, sA, LDSA, INFO)
\end{lstlisting}
and the ones for going from single to double are {\tt clat2z}
and {\tt slat2z}, respectively.

These routines, following LAPACK's interfaces, are also provided in
MAGMA for GPUs. MAGMA also adds conversion from single to 
half precision (FP32 to FP16) for general matrices:
\begin{lstlisting}[language=C++]
     slag2h(M, N, sA, LDA, hA, LDHA, INFO, QUEUE)
\end{lstlisting}
and the corresponding {\tt hlag2s}. These routines are well optimized
for NVIDIA GPUs, and also supported for AMD GPUs (through hipMAGMA).
MAGMA also provides the batched equivalent for batches of conversions.

A more specialized for mixed-precision calculations library may 
have to support a more complete set of data conversion routines, 
e.g., for arrays, strided arrays, tensors, etc.,
and more combinations of formats, including user/application defined.
For example, some mixed-precision FFT algorithms (see
Section~\ref{sec:fft})
use dynamic "splitting" of a high precision array (e.g., FP32) into
two lower-precision (e.g., FP16) arrays. See also Section~\ref{sec:format_decoupling} for further discussion and 
extensions on formats.

\subsection{Data compression}\label{sec:compress}
Data compression for reducing communications is another component
needed for the development of mixed-precision algorithms. 
The simplest form that we consider is the casting, as discussed
in Section~\ref{subs:casting}. This is an example of a lossy compression.
Casting from FP64 to FP32 for example, leads directly to a loss of 
about 8 decimal digits of accuracy, but reduces the data size by
a factor of two. Casting has been used to accelerate the FP64 solvers
in MAGMA up to $4\times$ using the mixed-precision iterative 
refinement 
techniques~\cite{hazw18,haidar2017investigating,haidar2018harnessing}
and we use it as benchmark to evaluate the potential of using other
compression algorithms.

We evaluated for example the possibility to use ZFP 
compression. ZFP provides lossy compression 
algorithms, where the compression mode can be specified by the 
user as either fixed rate, fixed accuracy, or fixed precision
~\cite{zfpv0.5}. Analysis for the round-off error introduced by 
ZFP in compressing floating-point data is presented
in~\cite{diffenderfer2018error}.
The values in
this experiment are taken random numbers and the compression 
specified is $4\times$. Note that compared to casting, the
compression rate is as casting to FP16, but the accuracy is
comparable to casting to FP32. These results make it feasible
to use tools like ZFP to accelerate memory-bound codes, 
e.g., like FFT (see Section~\ref{sec:fft}), up to $4\times$ 
while loosing about 8 decimal digits of accuracy.

\subsubsection{Mixed-precision MPI}\label{sec:mpmpi}
Of interest is MPI extension that fuses subsequent data conversions
with the MPI communications.
The conversion must be user specified and includes casting or
other data compression or conversion mechanisms, where a single 
MPI call will convert the input  
data as specified, send the converted data, and the corresponding MPI 
call will receive and convert the result again, as specified by 
the user. 
Our MPI collaborators have developed preliminary mixed-precision MPI 
for All2All and P2P communication using casting. The results show 
that asymptotically, for large enough data, the MPI communications 
can be accelerated proportional to the data compression, i.e., the
conversion is negligible. The implementations are for CPUs, as well
as GPUs using GPU-direct communications. 

\begin{figure}[tb]
\centering
\includegraphics[width=0.7\linewidth,height=7cm]{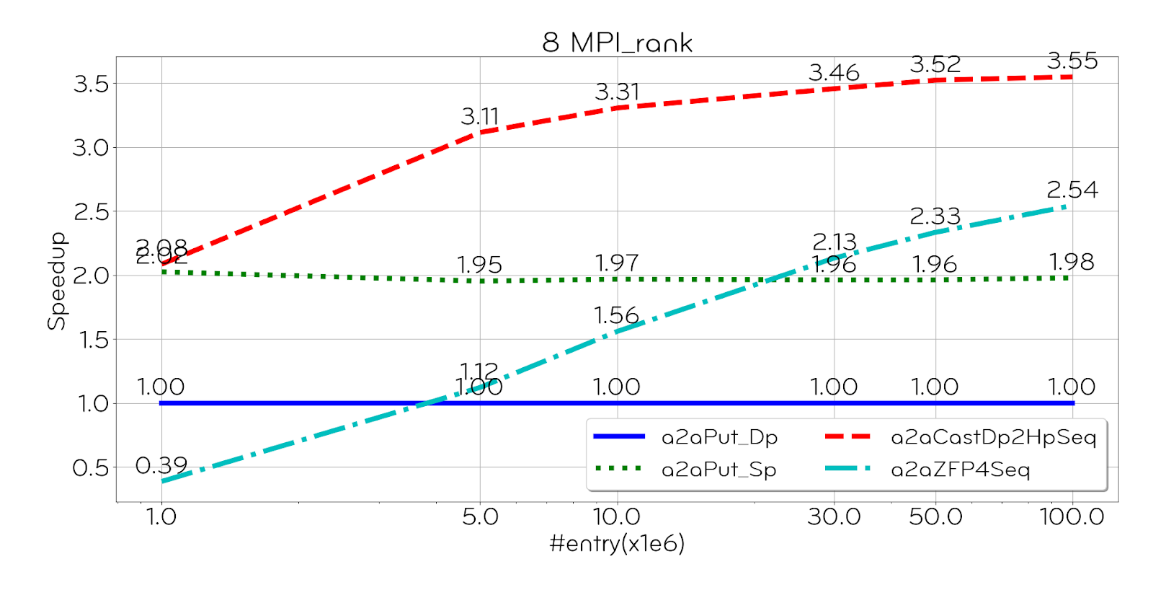}
\caption{Speedup of All2All by $4\times$ compression using
         cast (red) vs. $2\times$ compression using cast
         (dotted green) vs. $4\times$ compression using ZFP 
          on Nvidia V100 GPUs.}
\label{fig:zfpmpi}
\end{figure}

Our preliminary results can also use ZFP to compress the data.
Figure~\ref{fig:zfpmpi} illustrates an acceleration result
for All2All in FP64 (marked as base, i.e., 
the acceleration of 1 line).

Note that here we use ZFP and manage to accelerate the MPI 
communication more than $2\times$
while loosing only about $8$ decimal digits of accuracy, i.e., 
we achieve an acceleration that outperforms the corresponding
version that uses cast to FP32 (while having the same accuracy).
The data sending itself is accelerated
close to $4\times$ as it is compressed $4\times$, but the 
overall acceleration drops when adding the
cost for the data compression and decompression.
This means that acceleration theoretically still can go up 
to about $4\times$ in implementations where the GPU work on 
compression and decompression is pipelined and overlapped with 
the communication.

\subsection{Approximate Fast Fourier Transforms}\label{sec:fft}

One application of the mixed-precision technologies described
in this Section is the acceleration of multidimentional FFTs
through mixed-precision algorithms.
We found that more than dozen of ECP applications use FFTs in their codes, e.g., including LAMMPS, HACC, ExaAM, and applications from 
the Copa co-design center~\cite{fft-ecp}. ECP applications that require FFT-based solvers suffer from the lack of fast and scalable 3D FFT routines for distributed-heterogeneous parallel systems as the ones projected 
for the upcoming exascale computing systems, and some of the applications have indicated interest in exploring the use of
approximate FFTs that trade some loss of accuracy for increase in
performance.

\begin{figure}[tb]
\centering
\hspace{-0.3in}
\includegraphics[width=0.63\linewidth]{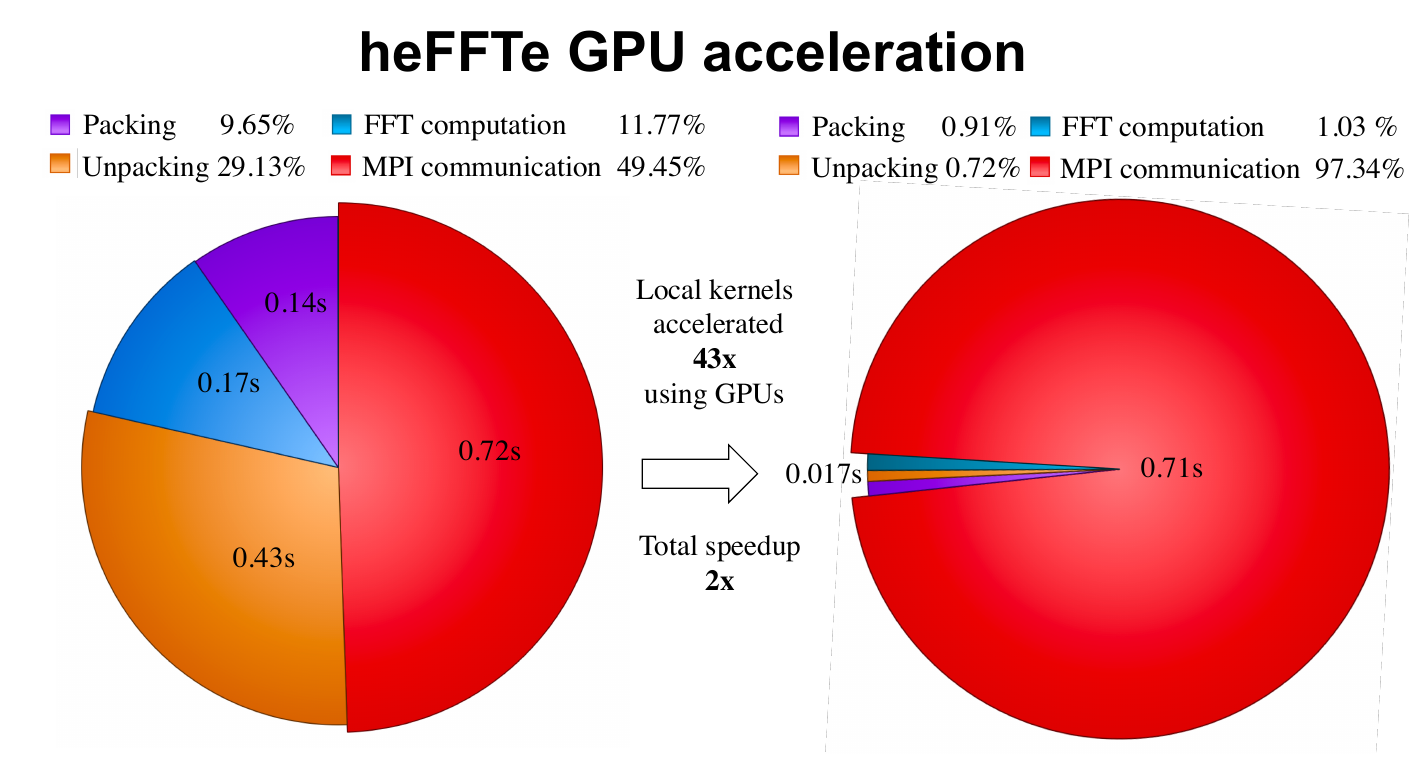}
\hspace{-0.1in}
\includegraphics[width=0.4\linewidth]{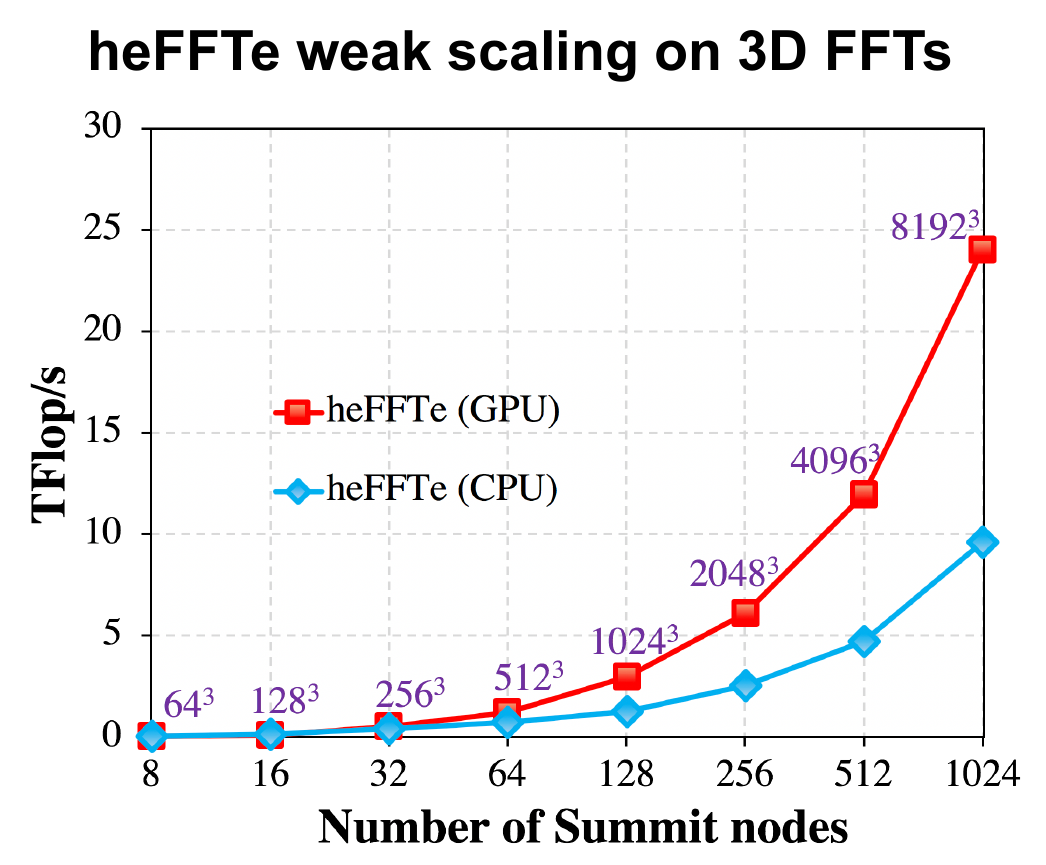}
\caption{{\bf Left}: heFFTe acceleration using GPUs vs. CPUs
         of $1024^3$ FFT on 
         4 Summit nodes. Note than nodal computations are accelerated 
         $43\times$ using GPUs. {\bf Right}: heFFTe weak scalability 
         with sizes indicated on the graph on up to 1024 nodes 
         (×6 V100 GPUs; 
         double complex arithmetic; starting and ending with bricks; performance assumes $5 N^3 log_2 N^3$ flops).}
\label{fig:heffte}
\end{figure}

To address the above needs for high-performance
scalable FFTs on large-scale GPU systems, we developed and released 
the {\it Highly Efficient FFTs for Exascale} ({\bf heFFTe}) 
library~\cite{heffte,fftmpi,gpudirectfft}. 
heFFTe v0.2 features very good weak, as well as 
strong, scalability and performance that is close to 90\% of 
the roofline peak (see Figure~\ref{fig:heffte}). However, 
after accelerating the local/nodal computations of about $43\times$
using GPUs (vs CPUs), the main bottleneck becomes the MPI communications. Currently, on typical FFT problems the GPUs can be 
used only about 1\% of the time, i.e., the GPUs are free to be used 
for other computations 99\% of the time, while the rest 99\% of the time is spent in MPI communications. Thus, any acceleration in the MPI
communications would translate into the same acceleration of the overall
FFT computation. 

\subsubsection{Approximate FFTs with accuracy-for-speed tradeoff}
One idea to accelerate FFTs using mixed-precision is 
through a tradeoff of accuracy for speed. Here we reduce
the communication volume by compressing the data, e.g.,
by casting or compression as outlined in Section~\ref{sec:compress}.

Multidimensional FFTs use All2All-type of MPI communications,
where first data is packed (locally on each GPU, using and
benefiting from GPUs' high bandwidth), next is the MPI communication,
and finally data is unpacked locally on each GPU~\cite{fft-ecp,heffte}.
As packing and unpacking are memory bound and involve going through
the data once, the addition of casting or other type of compression
can be fused with the packing/unpacking and thus significantly
remove its overhead. This idea is explored through the use of 
mixed-precision MPI that fuses all these operations 
(as in Section~\ref{sec:mpmpi}). Quantification of the speedups
obtained vs. the reduction in accuracy is as illustrated in Figure~\ref{fig:zfpmpi}.

Current work is on overlapping the use of the GPUs for 
compression/decompression with the MPI communications through
pipelining. This is possible as FFTs are memory bound and 
GPUs can be free to do other computations up to 99\% of the 
time, e.g., as benchmarked for the Summit hardware configuration. 

\subsubsection{Accuracy control}
Accuracy requirements are mainly application dependent and
can be controlled through specifying a desired compression
rate. 
The accuracy of the resulting
mixed-precision FFTs will be higher than the corresponding FFT
using "low" precision for both storage and computation, while
retaining the same performance. 
Moreover, going to the extreme, 
it is possible to tune the mixed-precision FFTs
to be of the same accuracy, e.g., as FP64 FFTs. For example,
this can be done through evaluating the loss of decimal digits 
in local computations due to round-off and use an appropriate
data compression rate, so that only the valid digits get to be 
communicated.

\subsubsection{Dynamic splitting}
Another idea in accelerating multidimensional FFTs is that
instead of compression, the high-precision FFT data can be 
dynamically split into two scaled data sets of lower-precision
data, apply the FFT transformations on the two sets in parallel
and combine the result at the end. This was explored in the context
of FP32 data that gets split into two FT16 sets in order to apply
fast GPU Tensor Cores computations on the FP16 data~\cite{sorna18,sornaSC18}, e.g.,
by going to a small radix, e.g., 4, where the FFT matrix can be
represented exactly (even in FP16) and benefit from Tensor Cores accelerated HGEMMs.
An extension of this idea can take into account that the 
FFT matrix is never assembled, except for a small radix matrix
in order to apply it on the data as GEMM. Thus, without much 
computational overhead (because of the memory-bound nature of FFT)
one can use higher precision Fourier matrix and computations 
(than the precision that stores the data) to accelerate the 
entire FFT computation.

%% file: sparse_factorization.tex
\newcommand{\xslnote}[1]{{\bf\color{red}[ #1 -- SHERRY ]}}
\newcommand{\precision}{\varepsilon}
\newcommand{\workingprec}{\precision_w}
\newcommand{\residualprec}{\precision_r}
\newcommand{\xprec}{\precision_x}
\newcommand{\iterind}[2]{#1^{(#2)}}
\newcommand{\truesoln}{x}
\newcommand{\soln}[1]{\iterind{\truesoln}{#1}}
\newcommand{\res}[1]{\iterind{r}{#1}}
\newcommand{\dx}[1]{\iterind{d\truesoln}{#1}}
    
\section{Multiprecision Sparse Factorizations}
\label{sec:sparse-direct}

Direct methods for sparse systems can also benefit from using lower precision formats.
The idea is to perform the expensive
calculations in lower precision, taking advantage of the faster speed often provided
by hardware. Then, some cheaper ``fixup''
algorithm is employed to recover the accuracy at a higher precision.
Sparse factorizations, such as sparse LU and QR factorizations,
are most often used to construct sparse direct solvers.
Two related but orthogonal research directions can be taken here.
The first is about the factorizations themselves, and the second
is in the context of direct solvers.

\subsection{Multiprecision sparse LU and QR}
Similar to dense LU and QR factorizations, a large fraction of the
computation lies in the Schur complement updates throughout the
elimination steps. In the dense case, much of the work in the Schur complement
update can be realized in terms of GEMM operations. However, in the sparse case,
each Schur complement update usually follows three steps:
1) gather the values from sparse data structures into contiguous memory,
2) perform GEMM operation,
3) scatter the output of GEMM into destination sparse data structures.

The benefit of using lower precision is two fold: for step 2), we can use
very fast lower precision vendor-provided GEMM functions, e.g. those utilizing NVIDIA's Tensor Cores. For the gather/scatter in steps 1) and 3), the
amount of data movement would be reduced.

For the dense case the main benefit comes from accelerated GEMM speed.
But in the sparse case, GEMM is only one part of the three steps above.
Furthermore, the dimensions of the GEMM kernel calls is generally smaller and of non-uniform
size throughout factorization. Therefore, the speed gain from GEMM alone is
limited. We will need to design new schemes to enhance overlap of GEMM
computation with gather/scatter operations.

\subsection{Multiprecision sparse direct solvers}
\label{sec:sparse-IR}

For the dense case, in \Cref{sec:denseiterref} we revisited the mixed precision iterative refinement (IR) algorithms with adaptive precision adjustment depending on convergence history.
The algorithms can
deliver high accuracy to the solution even when the expensive
LU and QR factorizations are done in lower precision.
We recall the IR algorithm using three precisions in~\Cref{alg:three_precision_IR_DS}~\cite{lin_itref06,lsq_itref09}.
This algorithm is already available in {\textbf xGERFSX} functions
in LAPACK.

The following three precisions are used:
\begin{itemize}
\item $\workingprec$ is the working precision used to store the input
  data $A$ and $b$. It is the lowest precision used in the solver, and is the desired precision for the output.
\item $\xprec$ is the precision used to store the computed solution
  $\soln{i}$. We require $\xprec \leq \workingprec$,
  possibly $\xprec \leq \workingprec^2$ if necessary
  for componentwise convergence.
\item $\residualprec$ is the precision used to compute the residuals
  $\res{i}$.  We usually have $\residualprec \ll \workingprec$,
  typically being at least twice the working precision
  ($\residualprec \leq \workingprec^2$).
\end{itemize}

\begin{algorithm}
    \caption{Three-precisions Iterative Refinement for Direct Solvers}
    \label{alg:three_precision_IR_DS}
    \begin{algorithmic}[1]
      \State Solve $A \soln{1} = b$ using the basic solution method (e.g., LU or QR)\;
        \Comment{($\workingprec$)}
    \State $i = 1$\; 
    \Repeat {
     \State $\res{i} \gets b - A\soln{i}$ \Comment{($\residualprec$)}
     \State Solve $A \, \dx{i+1} = \res{i}$ using the basic solution method\; \Comment{($\workingprec$)}
     \State Update $\soln{i+1} \gets \soln{i} + \dx{i+1}$\; \Comment{($\xprec$)}
     \State $i \gets i+1$\;
     \Until{$\soln{i}$ is ``accurate enough''}
     }\; \\
    \Return{$\soln{i}$ and error bounds}
    \end{algorithmic}
\end{algorithm}

With the above setup and adaptive adjustment of $\xprec$ and
$\residualprec$, 
the algorithm converges
with small normwise error and error bound if the normwise condition number of $A$ does not exceed $1/(\gamma(n)\workingprec)$.
Similarly, the algorithm converges with small componentwise error and error bound if the componentwise condition number of $A$ does not exceed $1/(\gamma(n)\workingprec)$.
Moreover, this IR procedure can return to the user the reliable error bounds both normwise and componentwise.
The error analysis in~\cite{lin_itref06} should all carry through
to the sparse cases.

The following are example configurations of the precisions:
\begin{itemize}
\item $\workingprec = 2^{-53}$ (IEEE-754 double precision),
  $\xprec = 2^{-53}$, $\residualprec = 2^{-106}$ (double-double)
\item $\workingprec = 2^{-16}$ (B-float),
  $\xprec = 2^{-24}$, $\residualprec = 2^{-53}$
\end{itemize}

Our plan is first to extend the above algorithm to the
sparse direct solvers SuperLU and STRUMPACK. 
While doing so, we will address the following open questions:
\begin{itemize}
\item When $\workingprec$ is bfloat16, the error analysis and error bounds may need be revisited. 
\item The relative cost of sparse LU/QR (lines 1 and 5) and sparse matvec (line 4) is different from the dense counter part. For a typical 3D PDE discretized problem, the respective costs are $O(n^2)$ and $O(n^{4/3})$. Thus, the ratio between "expensive" and "cheap" is smaller than the dense case. We need to be more mindful with the higher precision calculations.
\end{itemize}

%% file: krylov.tex
\section{Multiprecision efforts in Krylov solver technology}

The scope of our review includes both Lanczos-based (short-term recurrence) and Arnoldi-based (long-term recurrence) methods and the associated methods for solving linear systems of equations $Ax=b$. In the context of long-term recurrence methods, we consider the
Arnoldi-QR algorithm with the modified Gram-Schmidt implementation of the 
Generalized Minimum Residual (GMRES) Krylov subspace method for iteratively solving
linear systems of equations. The emphasis here is to examine the approaches
employed to date that incorporate mixed-precision floating point arithmetic to
speed-up computations, and yet retain some or all of the numerical properties
of the original algorithms in full double precision arithmetic (i.e.
representation error and loss of orthogonality).

\subsection{Lanczos-CG}

\subsubsection{Theoretical Results}
We first summarize very briefly the most well-known results on the finite precision behavior of Lanczos and CG methods, and discuss how such results could potentially be extended to the mixed precision case and existing progress in this area. We note that there is a huge literature on the finite precision behavior of Lanczos-based methods which we cannot hope to fully describe here. For a more thorough account and historical references, we point the reader to the manuscript of Meurant and Strako{\v s}~\cite{Meurant2006}. 

Fundamental relations dealing with the loss of orthogonality and other important quantities in finite precision Lanczos have been derived by Chris Paige~\cite{Paige1980}. These results were subsequently used by Anne Greenbaum to prove backward stability-like results for the CG method~\cite{Greenbaum89}; namely, Greenbaum showed that CG in finite precision can be seen as exact CG run on a larger linear system, in which the coefficient matrix has eigenvalues in tight clusters around the eigenvalues of the original matrix (where the diameter of these clusters  depends on properties of the matrix and the machine precision). Greenbaum also proved fundamental results on the maximum attainable accuracy in finite precision in what she calls ``recursively computed residual methods'', which includes CG, BICG, BICGSTAB, and other Lanczos-based methods~\cite{Greenbaum97}. The results of Paige and Greenbaum have also been extended to $s$-step Lanczos/CG variants in~\cite{Carson2015}, where it is shown that $s$-step Lanczos in finite precision behaves like classical Lanczos run in a lower ``effective'' precision (where this ``effective'' precision depends on the conditioning of the polynomials used to generate the $s$-step bases). 
We believe that these existing results can be extended to the mixed precision case; in Paige's analysis~\cite{Paige1980}, he first defines an $\eps_0$ quantity that is
used for errors in inner products and an $\eps_1$ quantity that comes from
errors in matrix-vector products, but then these quantities are combined in later theorems in order to simplify the analysis. It is possible to expand upon his analysis and keep these two quantities separate; such results could also then be interpreted in the framework of Greenbaum~\cite{Greenbaum89}.  

Existing results in the area of mixed precision Lanczos-based methods are contained within the work on ``inexact Krylov subspace methods'', which also applies to Arnoldi-based methods; see, e.g., the manuscripts of Simoncini and Szyld~\cite{Simoncini03}, and van den Eshof and Sleijpen~\cite{Eshof2004}. Within such frameworks, it is assumed that the 
matrix-vector products are computed with some bounded perturbation (which can change in each iteration) and all other computation is exact. These methods were motivated by improving performance in applications where the matrix-vector products dominate the cost of the computation, e.g., when the matrix is dense or the application of $A$ involves solving a linear system. Many theoretical results on ``inexact Krylov subspace methods'', mostly focused on the maximum attainable accuracy, have 
been proved in the literature. A surprising result is that the inexactness in the matrix-vector products can be permitted to grow in norm as the iterations progress at a rate proportional to the inverse of the residual norm without affecting the maximum attainable accuracy. However, a crucial practical question is whether inexactness will affect the convergence behavior \emph{before} the attainable accuracy is reached; this is entirely possible in the case of short-term recurrence methods such as CG and has not been well-studied theoretically. 

\subsubsection{Practical Applications}
We briefly mention works which make use of mixed precision Krylov subspace methods in practical applications, focusing on performance rather than on theoretical results.

One instance of this is in the work of Clark et al.~\cite{Clark2010}, which uses mixed precision CG and BICGSTAB methods implementing the ``reliable update'' strategy of Sleijpen and van der Vorst~\cite{Sleijpen1996} within a Lattice QCD application run on GPUs. The idea behind the ``reliable update'' strategy is that the true residual is computed and used to replace the recursively updated residual in select iterations, thus improving the attainable accuracy; this is done in conjunction with batched updates to the solution vector. By using higher (double) precision only in the true residual computations and group updates (and single or half precision for the rest of the computation), the authors claim they are able to achieve full double precision accuracy. This deserves further theoretical study, which we believe can be achieved by extending the results in~\cite{Sleijpen1996} and the related work of van der Vorst and Ye~\cite{vanderVorst2000} to the mixed precision setting.

\subsection{Arnoldi-$QR$ MGS-GMRES}

For MGS-GMRES the mixed precision work by Gratton et. al.~\cite{Gratton20} is the most recent
and appropriate - and in particular the loss-of-orthogonality relations due
to Bj\"orck~\cite{bjorck67a} and Paige~\cite{Paige1980}, later refined by Paige, Rozlo\v{z}nik and
Strako\v{s}~\cite{Paige2006}, are employed in order to provide tolerances for mixed single--double
computations.  MGS-GMRES convergence stalls (the norm-wise relative backward error approaches
$\eps$) when linear independence of the Krylov vectors is lost, and this is 
signaled by Paige's $S$ matrix norm $\|S\|_2 = 1$. The $S$ matrix \cite{Paige18} is derived 
from the lower triangular $T$ matrix appearing in the rounding error analyses by 
Giraud et.~al.~\cite{Giraud2004}. 

To summarize, the Gratton et. al.~\cite{Gratton20} paper postulates starting from the
Arnoldi-QR algorithm using the modified Gram-Schmidt algorithm and
employing exact arithmetic in the MGS-GMRES iterative solver.  The Arnoldi-QR
algorithm applied to a non-symmetric matrix $A$ produces the matrix
factorization, with loss of orthogonality $F_k$
\begin{equation}
AV_k = V_{k+1}\:H_k, \quad V_{k+1}^TV_{k+1} = I + F_k
\end{equation}
They next introduce inexact (e.g. single precision) inner products - this
directly relates to the loss-of-orthogonality relations for the $A = QR$ factorization
produced by MGS. The resulting loss of orthogonality, as measured by $\|I - Q^TQ \|_2$,
grows as ${\cal O}(\eps)\kappa(A)$ as was derived by Bj\"orck~\cite{bjorck67a} and ${\cal
O}(\eps)\kappa ( [\: r_0,\: AV_k\:] )$ for Arnoldi-QR - which is described by
Paige, Rozlo{\v z}n{\' i}k and Strako{\v s}~\cite{Paige2002, Paige2006} and related work.  
The inexact inner products are given by
\begin{equation}
h_{ij} = v_i^Tw_j + \eta_{ij}
\end{equation}
where $h_{ij}$ are elements of the Hessenberg matrix $H_k$, and the Arnoldi-QR algorithm
produces a $QR$ factorization of the matrix
\begin{equation}
\left[
\begin{array}{cc}
r_0,& AV_k
\end{array}
\right]
= V_{k+1}
\left[
\begin{array}{cc}
\beta\:e_1,& H_k
\end{array}
\right], 
\end{equation}
The loss of orthogonality relations for $F_k$ are given below,
where the matrix $U$ is strictly upper triangular
\begin{equation}
F_k = \bar{U}_k + \bar{U}_k^T, \quad
U_k = \left[
\begin{array}{ccc}
v_1^Tv_2 & \cdots & v_1^Tv_{k+1} \\
         & \ddots &              \\
         &        & v_k^Tv_{k+1} 
\end{array}
\right] 
\end{equation}
Define the matrices,
\begin{equation}
N_k = \left[
\begin{array}{ccc}
\eta_{11} & \cdots & \eta_{1k} \\
         & \ddots &              \\
         &        & \eta_{kk} 
\end{array}
\right], \quad
R_k = \left[
\begin{array}{ccc}
h_{21} & \cdots & h_{2k} \\
       & \ddots &              \\
       &        & h_{k+1,k} 
\end{array}
\right]
\end{equation}
The loss of orthogonality relation derived by Bj\"orck \cite{bjorck67a}, for the $A=QR$ factorization
via the modified Gram-Schmidt algorithm can be applied to the Arnoldi-QR algorithm to obtain
\begin{equation}
N_k = -
\left[
\begin{array}{cc}
0,& U_k
\end{array}
\right]\: H_k = 
- U_kR_k
\end{equation}
The complete loss of orthogonality (linear independence) of the Krylov vectors in
MGS-GMRES signals the minimum error is achieved and GMRES then stalls or really
can go no further than when the norm-wise relative backward error reaches ${\cal O}(\eps)$.
Gratton et al.~\cite{Gratton20} show how to maintain sufficient orthogonality in order
to achieve a desired relative residual error level - by switching the inner products
from double to single at certain tolerance levels and combine this with
inexact matrix-vector products as in van den Eshof and Sleijpen~\cite{Eshof2004} and 
Simoncini and Szyld~\cite{Simoncini03}.

In practice, the restarted variant of GMRES is often employed to reduce memory requirements.
The algorithm produces both implicit and explicit residuals. Thus, we might ask whether either 
can be performed in reduced precision.
The work described herein on iterative refinement by Nick Higham and Erin Carson for mixed precision 
can be applied to analyse the convergence of restarted GMRES$(m)$, assuming 
a fixed number of iterations - because restarted GMRES is just iterative refinement 
with GMRES as the solver for the correction term. However, a more detailed analysis
with experiments has yet to be performed. We are fairly certain that
the residual computations must be performed in higher precision in 
order to achieve a norm-wise backward error close to double precision machine round-off.

\subsection{Alterative Approaches}

Although somewhat outside the scope of this review, we can demonstrate that it is
possible to modify the Gratton et al.~\cite{Gratton20} analysis based on the inverse
compact $WY$ form of the MGS algorithm, introduced by \v{S}wirydowicz et al.~\cite{Swirydowicz19}. 
Rather than treat all of the inner products in the MGS-GMRES algorithm equally,
consider the strictly upper triangular matrix $U = L^T$ from the loss of orthogonality relations.
We introduce single precision $L_{:,j-1} = Q_{j-1}^Tq_{j-1}$ and double precision
triangular solve $r = (I + L_{j-1})^{-1} Q_{j-1}^Ta$ to update $R$ - as this
would directly employ the forward error analysis of Higham~\cite{Higham89}.  The
former affects the loss of orthogonality, whereas the latter affects the
representation error for $QR$ - but then also for Arnoldi-QR.  This could allow
more (or most) of the inner products to be computed in single precision.

\begin{figure}[p]
\centering
\includegraphics[width=0.7\textwidth]{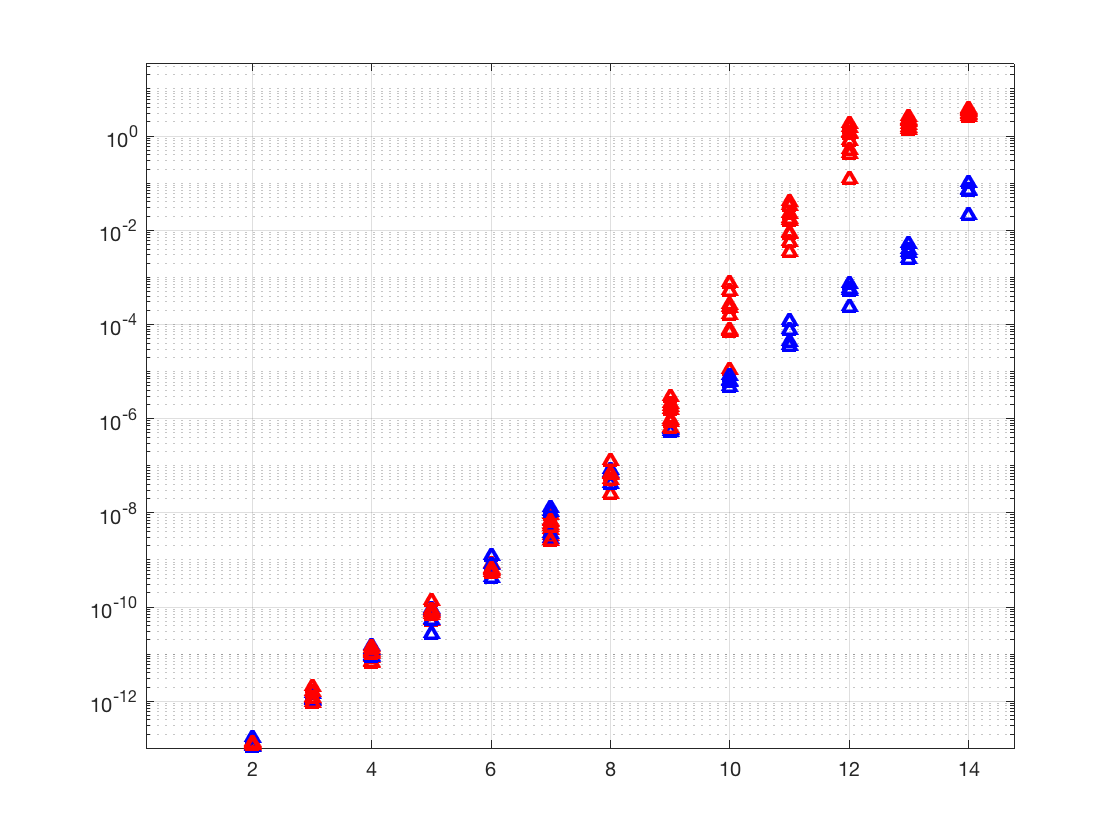}
\caption{Loss of Orthogonality for Mixed Single-Double MGS Algorithm}
\label{fig:krylov1}
\end{figure}

Evidence for maintaining orthogonality is provided in~\Cref{fig:krylov1}, with $\|I - Q^TQ\|$ 
plotted for $A=QR$ using the inner products in standard MGS (blue) in double precision
versus the inverse compact WY MGS (red) with $Q_{j-1}^Tq_{j-1}$ in single precision 
(simulated in MATLAB) - and we observe at least the same or slightly higher error
levels. The $x$-axis is log condition number for randomly generated matrices. 
The lower triangular solve is computed in double precision.

Barlow ~\cite{Barlow19} contains similar if not  the  same  algorithm  formulations
in  block form.  His work is related to Bj\"orck's 1994 paper \cite[Section 7]{Bjorck94} 
which derives 
the triangular matrix $T$ using a recursive form for MGS, and which is referred to as
a ``compact WY" representation in the literature.  While Bj\"orck used a lower triangular matrix
for the compact WY form of MGS, Malard and Paige \cite{Malard94}  derived  the  upper
triangular form, also employed by Barlow, which reverses the order of elementary projectors.  
The latter is  unstable in  that  a  backward recurrence leads to ${\cal O}(\eps)\kappa^2(A)$ loss of orthogonality. 
An interesting observation from Julien Langou is that the upper triangular form is less stable than the lower
triangular (even though the backward-forward algorithm results in re-orthogonalization; 
see the algorithm in Leon, Bj\"orck, Gander~\cite{Leon10}).

Barlow \cite{Barlow19} employs the Householder compact WY representation of reflectors and also 
refers to the work of Chiara Puglisi \cite{Puglisi92} -- discussed in Joffrain et al. \cite{Joffrain06} -- 
and this is referred to as the ``inverse compact WY" representation of Householder; this
originally comes from Walker's work on Householder GMRES~\cite{Walker88}.  Barlow then
extends this approach to the block compact WY form of MGS; see also the technical report
by Sun \cite{Sun96}.  The contribution by \v{S}wirydowicz et al.~\cite{Swirydowicz19} 
was to note that there exists an inverse compact WY representation for MGS - having the projector 
\[
P^{IM} = I - Q_{j-1} T^{IM} Q_{j-1}^T = I - Q_{j-1} (I + L_{j-1})^{-1} Q_{j-1}^T
\]
and to ``lag" the norm $\|q_{j-1}\|_2$ so that these can be computed in one
global reduction. Barlow \cite{Barlow19} makes this connection for blocks (and in effect
this is given in his equation (3.10)) and references Puglisi \cite{Puglisi92}. 

Bj\"orck and Paige \cite{Bjorck92} made the link between Householder and MGS based on
the observation made by Sheffield. Paige defines this to be augmentation and
Gratton et al. \cite{Gratton20} also references this work.  Paige has also recently
extended these augmentation ideas to Lanczos.  The $T$ matrix appears in
Paige's work with W\"{u}lling~\cite{Paige14} and then later in \cite{Paige18} to derive the loss of
orthogonality matrix $S = (I + L_{j-1}^T)^{-1} L_{j-1}^T$. This also appears in
the work of Giraud, Gratton and Langou~\cite{Giraud2004}; Langou also worked with Barlow
and Smoktunowicz \cite{Smok06} on the Pythagorean trick to reduce cancellation error
in the computation of vector norms and a Cholesky-like form of classical Gram-Schmidt (CGS).

In order to combine single-double floating-point operations in MGS-GMRES, at first it appears that we
could store the $T$ matrix in single precision - but then we would still have to form $Q_{j-1}^Ta$, 
and store $Q_{j-1}$ in double precision. By examining the cost trade-offs a bit further, 
we can instead use a form of re-orthogonalization based on a backward-forward solver recurrence
\[ 
T = (I + L_{j-1}^T )^{-1}\: ( I + L_{j-1} )^{-1}
\] 
and our initial computational results demonstrate this works well, driving
the relative residual to ${\cal O}(\eps)$ in double, with orthogonality
maintained to ${\cal O}(\eps)$ in single.

The representation error (backwards error) for $A + E = QR$ computed by MGS, is
not affected by single precision inner products - and remains ${\cal O}(\eps)$. 
We are not aware of whether or not this was previously known.

\begin{figure}[p]
\centering
\includegraphics[width=0.7\textwidth]{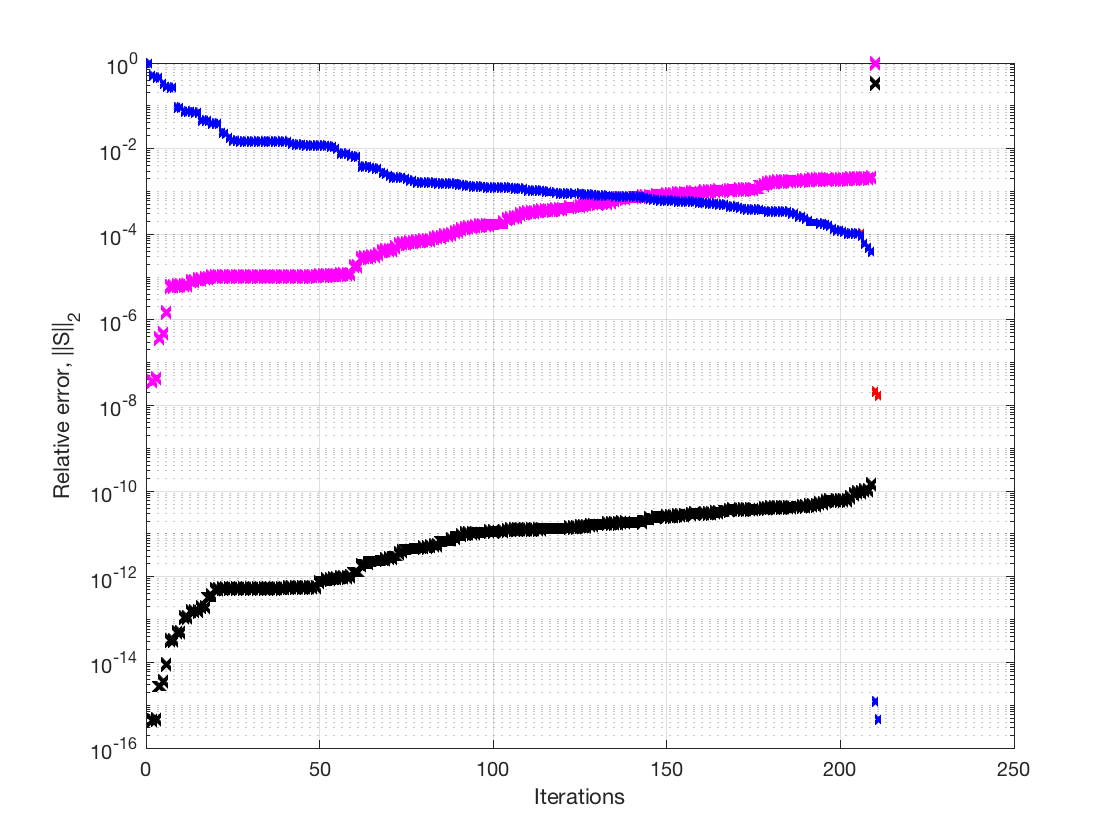}
\caption{GMRES residuals and loss of orthogonality $\|S\|_2$ for impcol\_e matrix}
\label{fig:krylov2}
\end{figure}

%% file: preconditioning.tex
\section{Multiprecision Preconditioners}
In the iterative solution process of large sparse systems, preconditioners are an important
building block facilitating satisfactory convergence.
The concept of preconditioning is to turn  an ill-conditioned liner system $Ax=b$ into a (left-) preconditioned system
$MAx=Mb$ ($AMy=b$, $x=My$ for right-preconditioning), which allows for faster
convergence of the iterative solver.
The convergence characteristics typically depend on the conditioning of the target system.
For an ill-conditioned $A$, the preconditioner is also required to be ill-conditioned. Otherwise, the preconditioner can not be expected to
improve the conditioning of the problem or the convergence of the iterative solver.
In that respect, the preconditioner basically tries to approximate the
inverse of the system matrix. Obviously, if the preconditioner is the
exact inverse, the solution is readily available. However, computing the
exact inverse is prohibitively expensive, and in most cases, the
preconditioner is just a rough approximation of the system matrix
inverse.  As a consequence, it is natural to question the need for using
high precision for a preconditioner that is inherently carrying only
limited accuracy. Indeed, choosing a lower precision format for the
preconditioner is a valid strategy as long as the accuracy loss induced
by using a lower precision format neither impacts the preconditioner
accuracy nor its regularity. For example, Trilinos allows the use of low
precision preconditioners inside high precision iterative solvers,
see~\Cref{interoperability}, and the hypre team works on multigrid
methods running the first cycles in lower precision. However, the use of
lower precision in the preconditioner application results in different
rounding effects than when using high precision. Specifically, the
rounding effects make the preconditioner non-constant as the rounding
effects are not only larger than in high precision, but also depend on
the input data~\cite{anzt2019adaptive}. As a result, low precision
preconditioners can only be used to accelerate an iterative method that
can handle non-constant preconditioners, i.e., can converge even if the
preconditioner changes in-between iterations. For the Krylov subspace
solvers generating search directions orthogonal to the previous search
direction, a changing preconditioner requires an additional
orthogonalization of the preconditioned search direction against the
previous preconditioned search direction. The flexible Krylov solvers
(e.g. FGMRES, FCG) contain this additional orthogonalization and are
therefore slightly more expensive. At the same time, they 
do allow for using low precision preconditioners, which can compensate for the additinal cost.

An alternative workaround is to decouple the memory precision from the
arithmetic precision, see~\Cref{sec:format_decoupling}, and only store the
preconditioner in low precision but apply it in high
precision~\cite{anzt2019adaptive}. Running all arithmetic in high
precision keeps the preconditioner constant, and removes the need for the additional orthogonalization of the preconditioned search direction. 
On the other hand, decoupling memory
precision from arithmetic precision requires to convert in-between the
formats on-the-fly when reading data from main memory. Fortunately, most
iterative solvers and preconditioners are memory bound, and the
conversion can be hidden behind the memory
transfers. A production-ready implementation
of an adaptive precision block-Jacobi preconditioner decoupling memory
precision from arithmetic precision is available in the Ginkgo library,
see~\Cref{interoperability}.

%% file: format_decoupling.tex
\section{Multiprecision efforts decoupling the arithmetic format from the memory format}
\label{sec:format_decoupling}
Across the complete hardware technology foodchain, we are witnessing a widening gap between the compute power in terms of float point operations per second on the one side and the communication power in terms of memory bandwidth. 
In modern processor technology, retrieving values from main memory takes 
several orders of magnitude longer than performing arithmetic operations, and communicating between
distinct nodes of a cluster is again orders of magnitude slower than main memory access. 
In consequence more and more algorithms hit the memory wall -- and already today, virtually all 
applications inside the ECP ecosystem are memory bound on modern hardware architectures.
With no disruptive hardware changes on the horizon, we are facing a situation where all applications
suffer from the slow communication to main memory or in-between nodes.\\

\begin{figure}
    \centering
    \includegraphics[width=.7\textwidth]{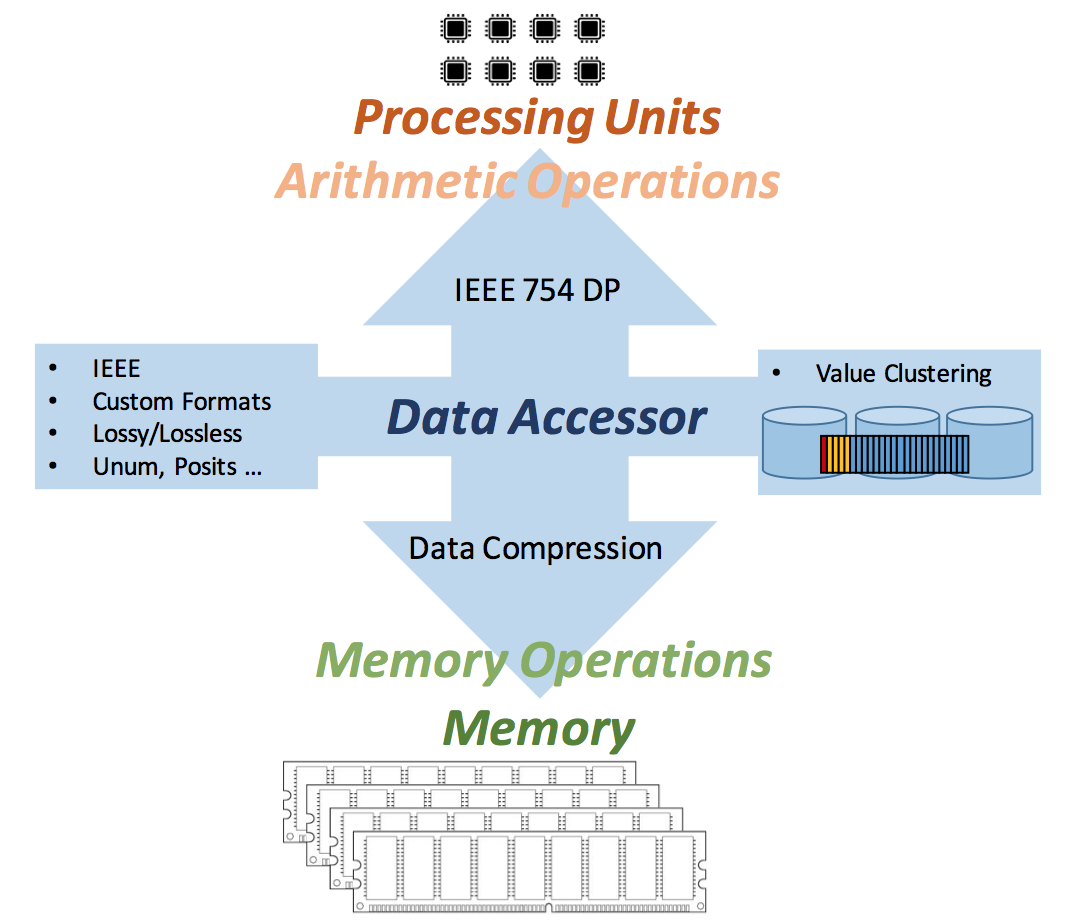}
    \caption{Accessor separating the memory format from the arithmetic format and realizing on-the-fly data conversion in each memory access.}
    \label{fig:accessor}
\end{figure}

A promising -- and maybe the only promising -- strategy to overcome this 
problem is to utilize the bandwidth capacity more carefully, reduce the 
communication volume and the number of communication points, and whenever 
possible, trade communication against computations. Specifically,
the idea is to radically decouple the memory precision from the arithmetic
precision, employ high precision only in the computations, and lower the precision 
as much as possible when accessing data in main memory or communicating with 
remote processors~\cite{anzt2019toward}. An important aspect in this context is
the design of a ``\textit{memory accessor}'' that converts data on the fly
between the IEEE high precision arithmetic format and the memory/communication format, see Figure~\ref{fig:accessor}.
Also, the memory/communication format does not necessarily have to be part of 
the IEEE standard, but can also be an arbitrary composition of sign, exponent, and 
significand bits~\cite{grutzmacher2019customized}, 
or even nonstandard formats like L. Gustafson's Unum format~\cite{unum}.
Obviously, there is a close link to the idea to complement the format separation
with compression techniques, like proposed in Section~\ref{sec:compression}.\\

\begin{figure}
    \centering
    \includegraphics[width=.5\textwidth]{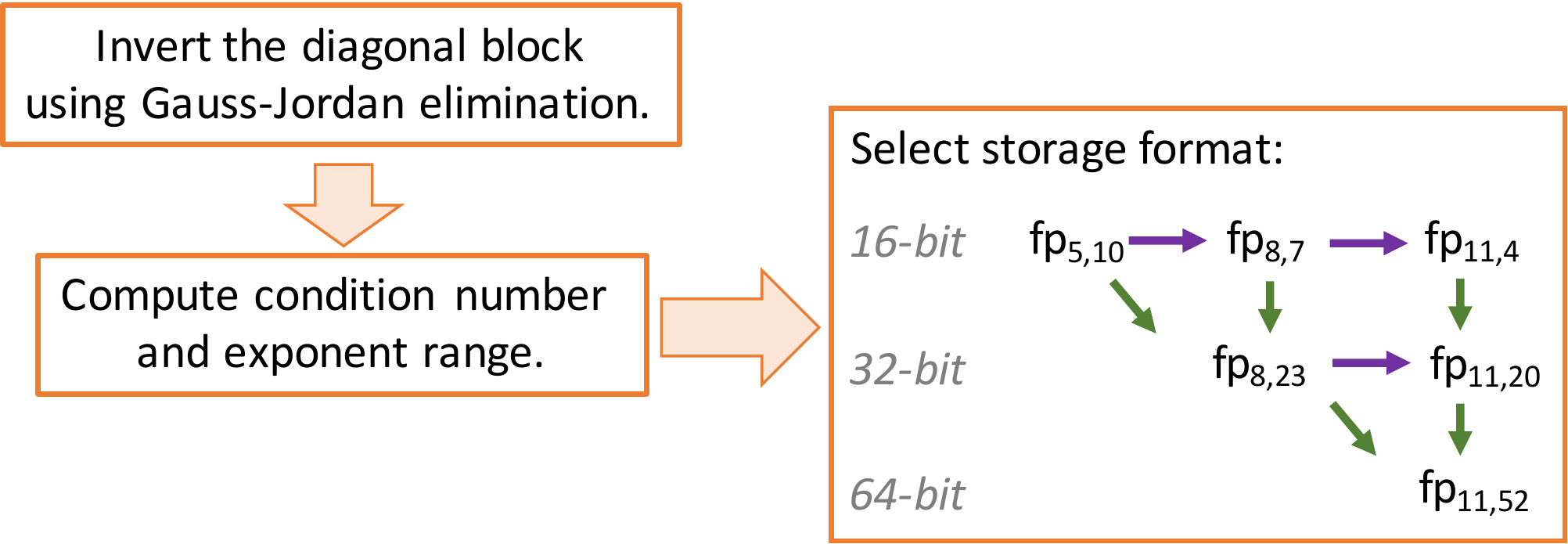}
    \caption{Storage format optimization for block-Jacobi: Starting from the most compact storage (left top), the format is extended in exponent bits to fit the data range (rightwards) and to preserve regularity (downwards) until both is satisfied.}
    \label{fig:adaptiveblockjacobistrategy}
\end{figure}
A proof-of-concept for the idea of decoupling arithmetic precision from memory
precision is the adaptive precision block-Jacobi preconditioner~\cite{anzt2019adaptive}
available in the Ginkgo sparse linear algebra library.
The idea here is to compute a block-Jacobi preconditioner in high precision, but then 
store the distinct inverted diagonal blocks in the lowest floating point precision 
format that avoids overflow and still preserves the regularity of the preconditioner, see Figure~\cref{fig:adaptiveblockjacobistrategy}.

\begin{figure}
    \centering
    \includegraphics[width=\textwidth]{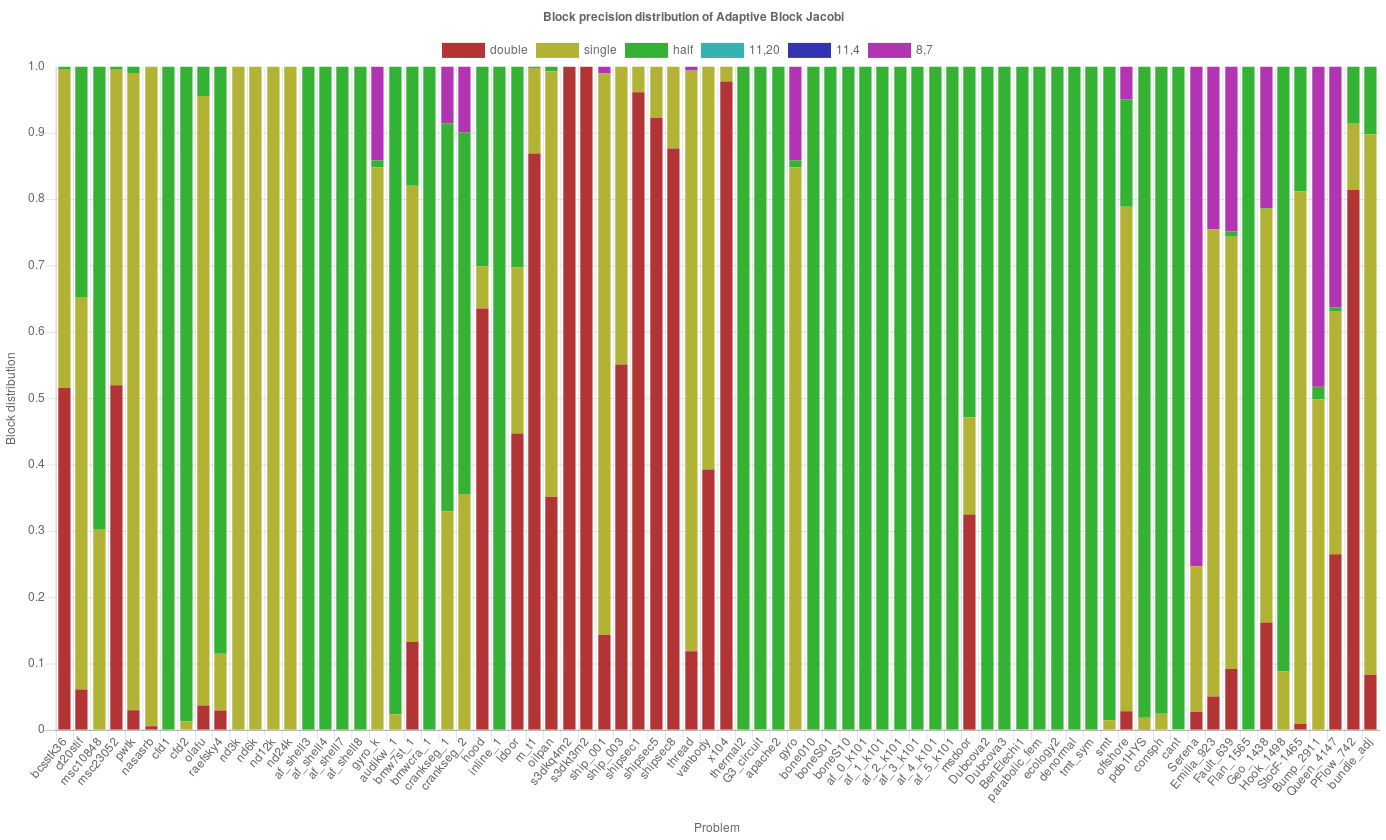}\\
    \includegraphics[width=\textwidth]{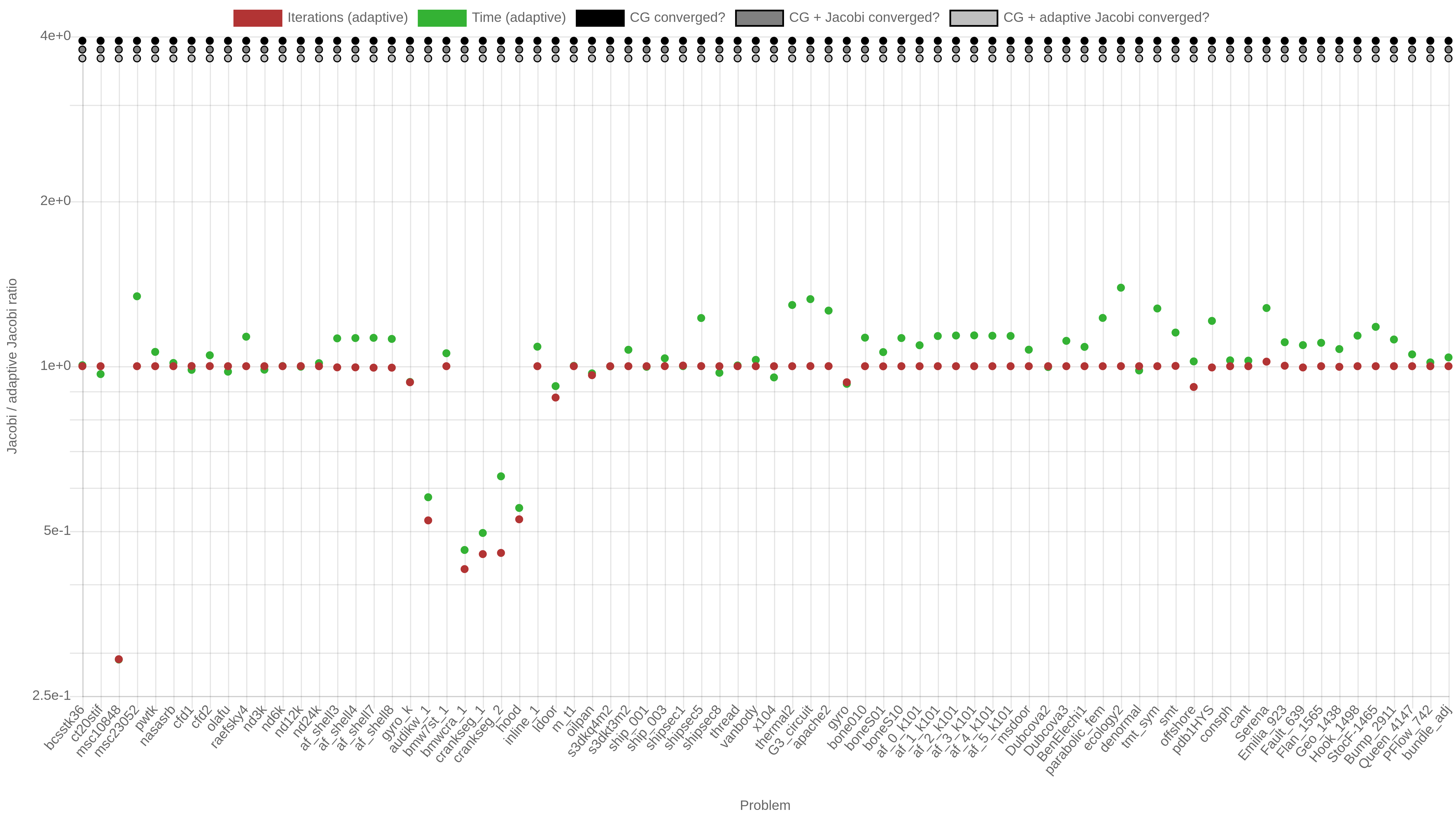}
    \caption{Top: Distribution of floating point formats among the distinct blocks when 
preserving 1 digit accuracy of each inverted diagonal block.
Each column represents one symmetric positive definite matrices of the Suite Sparse Matrix Collection. Bottom: Impact on the top-level CG solver solving the system-induced linear problem. For most systems, the convergence rate is unaffected by the use of a lower storage precision format, all preconditioner applications are faster, resulting in an average 20\% runtime reduction.}
    \label{fig:formatdistribution}
\end{figure}
This storage format is chosen for each diagonal block individually, respectively reflecting the characteristics.
Figure~\ref{fig:formatdistribution} (top) visualizes the distribution of formats when storing the inverted diagonal blocks of size 24 for symmetric positive definite matrices of the Suite Sparse Matrix Collection.
Obviously, converting to a lower format generally reduces the accuracy of the
linear operator, but as block-Jacobi preconditioners ignore all off-(block)diagonal entries,
they are typically only a rough approximation of the matrix inverse and therewith by design 
only have very limited accuracy. Experimental results reveal that the use of a 
lower precision format for storing the inverted diagonal blocks has in most cases only negligible 
effects on the preconditioner effectiveness and the outer solver convergence.
At the same time, storing the inverted diagonal blocks in lower precision reduces the
memory access volume in every preconditioner application, therewith accelerating
the bandwidth bound iterative solution process, see Figure~\ref{fig:formatdistribution}.
For the adaptive precision block-Jacobi preconditioner, is important that the accessor 
converts the inverted diagonal blocks back to the IEEE standard precision not only for
performance reasons -- leveraging the highly-optimized IEEE floating point arithmetic
of the processors -- but also for numeric reasons: Using working precision in the 
arithmetic operations of the precnditioner application preserves the preconditioner as a 
constant operator, applying a preconditioner in lower precision would result in a 
non-constant preconditioner and require the use of a (more expensive) 
flexible iterative solver~\cite{anzt2019adaptive}.

%% file: multigrid.tex
\subsection{Using different precision formats in Multigrid methods}
Multigrid methods are highly effective iterative methods. There are basically two types of multigrid methods: geometric multigrid methods (GMG) and algebraic multigrid methods (AMG). GMG requires actual grids on each level to generate its components, whereas AMG can be considered more like a ‘black box’ method, in that it can be given a matrix and right hand side and will generate the components for each level automatically using sensible heuristics. These methods are an interesting target for multiprecision treatment due to their different components which affect the overall algorithm in different ways.
GMG and AMG components combine smoothers, coarser grid, restriction and prolongation operators on each level. In addition, it is of interest to investigate changes in precision on different levels. Finally, GMG and AMG can be used as preconditioners to other solvers, i.e. there is potential to use lower precision across the whole preconditioner.
Historically, most work focused on the use of a lower precision GMG or AMG method as a preconditioner to a double precision solver. Additionally, there are attempts to apply ZFP~\cite{zfpv0.5} within MG or establish an error analysis framework for AMG.

Ljungkvist and Kronbichler\cite{LjKr2017,LjKr2019} successfully use mixed precision to solve the Laplace problem for different orders with a matrix-free geometric multigrid approach. Their solver infrastructure allows for using mixed-precision arithmetic that performs the multigrid V-cycle in single precision with an outer correction in double precision, increasing throughput by up to 83 percent.

Similarly, Glimberg et al \cite{Gletal2011} use a single precision multigrid to precondition a double precision defect correction scheme and solve the Laplace problem within a nonlinear water wave application on a GPU architecture. They achieve a speedup of up to 1.6 of the mixed precision version over the double precision version, a speedup of 1.9 for a purely single precision version. 

Yamagishi and Matsumura \cite{YaMa2016} also apply a single precision multigrid to a double precision conjugate gradient solver to the Poisson/Helmholtz problem within their non-hydrostatic ocean model. They report a speedup up to 2 for a single precision Matvec over a double precision one and improved overall times using this approach, however they compare the full application run only to their CPU version.

There are various publications that pursue the same strategy of using a single precision AMG preconditioner to a double precision solver. 

Emans and van de Meer \cite{EmvM2010}perform a careful analysis of the individual kernels of preconditioned Krylov solvers on multi-core CPUs, including sparse matrix-vector multiplications (SpMV) which make up a large portion of AMG. They also consider the effect of communication, where lower precision leads to smaller size messages, but latencies are still an issue, particularly on the coarsest levels of AMG.  They find that the use of mixed precision for the preconditioner barely affects convergence and therefore speedups for the kernels, which were between 1.1 and 1.5, can potentially carry over to the whole solver and lead to improvements of runtimes within CFD applications. 

Sumiyoshi et al \cite{Suetal2014} investigate AMG performance on a heterogeneous computer architecture with both CPUs and GPUs for isotropic and anisotropic Poisson problems. They consider smoothed aggregation AMG as a stand-alone solver. They carefully analyze different portions of the algorithm on five different architectures, including one multi-core CPU cluster. They report speedups between 1.2 and 1.6 on the GPU-CPU architectures for the mixed-precision implementation over the double precision version. These speedups are related to SpMV performance (between 1.6 and 1.8) on these architectures. However, the mixed-precision version was slightly slower on the CPU-only architecture, which achieved barely any improvement for the SpMV operations.

Richter et al \cite{Rietal2014} examine the performance of a single precision AMG solver (ML and PETSc) applied to a double precision PCG solver. They apply the method for an electrostatic simulation of the high voltage isolator on a GPU/CPU computer architecture. Their mixed precision version takes about 84 percent of the time of the double precision version.

An approach described in a presentation by Kate Clark \cite{Clark2019} takes the use of mixed precision even further to involve half precision. Clark and collaborators achieved good results using a double precision defect correction approach with a single precision Krylov solver and a half precision AMG preconditioner.

Another interesting related study by Fox and Kolasinski \cite{FoKo2019} examines the use of ZFP, a lossy compression algorithm, within multigrid. Due to the local structure of ZFP, ZFP can easily be integrated into numerical simulations without changing the underlying algorithms. However, since ZFP is a lossy algorithm, it will introduce some error, thus, it is important to understand if the error caused by ZFP overwhelms other traditional sources of error, such as discretization error. The study shows that for MG on a Poisson problem, applying ZFP to the approximation vector, can significantly decrease memory use and thus is expected to decrease runtimes, while the generated errors stay below discretization error. Since a hardware version of ZFP is not available yet, no actual runs were possible, however the results show good potential to use GMG and/or AMG as a preconditioner.

Currently, Tamstorf et al \cite{Taetal2020} appear to be the only ones who have investigated
the theory of multi-precision multigrid methods. Their original intent was to
improve the appearance of the movement of cloth within Disney movies, which
requires higher than FP64 accuracy. However, their theory
applies equally to decreased precision. They have created
a theoretical framework with rigorous proofs for a mixed-precision version
of multigrid for solving the algebraic equations that arise from
discretizing linear elliptic partial differential equations (PDEs).  The arising matrices 
being sparse and symmetric positive definite
enable the use of the so-called energy or $A$ norm to establish convergence
and error estimates. Bounds on the convergence
behavior of multigrid are developed and analyzed as a function of the matrix condition number. Both
theoretical and numerical results confirm that convergence
to the level of discretization accuracy can be achieved with mixed-precision
versions of V-cycles and full multigrid.   This framework is inspired
by the results of Carson and Higham \cite{carson2017new} but
ultimately provides tighter bounds for many PDEs. 
Tamstorf et al \cite{Taetal2020b} further extend their theoretical framework to include the quantization error.
They use the bounds to guide the choice of precision level in their progressive-precision multigrid scheme
by balancing quantization, algebraic and descretization errors. They show that while iterative refinement is susceptible to quantization errors during the
residual and update computation, the V-cycle used to compute the correction in each iteration is
much more resilient, and continues to work if the system matrices in the hierarchy become indefinite
due to quantization.


%% file: fft.tex

%% file: ml.tex
\section{Low precision and multiprecision technology for Machine Learning}

Modern \ac{hpc} hardware continues to experience an ongoing shift
towards supporting a variety reduced-precision formats for representing
floating-point numbers in order to offer a much increased performance
rate. However, portability is often of little concern as the hardware
tends to serve only a specific set of workloads that are of special
interest to the particular vendor.  The examples include Intel's Cascade
Lake Vector Neural Network Instructions (VNNI) and the recently
announced Xe platform for graphics cards, AMD's Radeon Instinct cards
(MI5, MI8, MI25, MI55, MI60) and NVIDIA's compute cards from the Pascal,
Volta, and Turing series. Finally, ARM included 16-bit floating point
(FP16) in its NEON vector unit specification VFP 8.2-A.  These
accelerators follow two types of specifications for 16-bit
floating-point format: IEEE-compliant FLOAT16 and extended-range
BFLOAT16.

At the same time, a new breed of accelerators take the use of reduced
precision to a new level as they target new machine learning workloads.
This new hardware includes  Cloud AI 100 by Qualcomm, Dot-Product Engine
by HPE, Eyeriss by MIT~\cite{DBLP:journals/corr/abs-1807-07928}, TPU by
Google~\cite{DBLP:journals/corr/JouppiYPPABBBBB17}, MAERI by Georgia Institute of Technology~\cite{kwon2018maeri} Deep Learning Boost
by Intel, CS-1 by Cerebras, and Zion by Facebook.

In general, the machine learning community has been more aggressive in evaluating multiple precision to the extent that even a 1-bit Stochastic Gradient Descent has been considered~\cite{seide20141}. The typical use case in machine learning is to use the training with 32-bit arithmetic and use different precision for the inference task. The quantization for the inference is supported in popular frameworks like TensorFlow~\cite{abadi2016tensorflow} and pyTorch~\cite{paszke2017automatic}. Quantization is the approach to store the tensors and compute on them using bitwidths lower than floating point bitwidths. Even in machine learning frameworks, the support for quantizations is limited to just the key functionality needed for a convolutional neural networks or recurrent neural networks with some limited hardware support. For example, pyTorch and TensorFlow supports 8-bit quantization for activation and weights. This allows using 8-bits for inference where the additional 2-4x performance is necessary. On the training front, it has been shown that 16-bit training is sufficient for certain tasks~\cite{gupta2015deep,courbariaux2014training}. The recent Gordon Bell winner demonstrated that lower-precison training can be used for scientific machine learning tasks as well~\cite{kurth2018exascale}.

The analogous effort to the work in deep learning to the examples of our interest in scientific computing involves training the network in
lower precision and performing inference in a higher
one~\cite{DBLP:journals/corr/GuptaAGN15,Gupta:2015:DLL:3045118.3045303}.
The compute imbalance between training and inference is even higher than
that of factorization and the subsequent iterative refinement. Another
difference is that in the context of neural network training, lowering
the precision may be incorporated into the model as a regularizer.

%% file: interoperability.tex
\section{Multiprecision capabilities of xSDK Math Libraries and Interoperability}
\label{interoperability}

\subsection{Ginkgo}
Ginkgo is a modern sparse
linear algebra library able to run on multi- and manycore architectures~\cite{anzt2020preparing}. The
library design is guided by combining ecosystem extensibility with heavy,
architecture-specific kernel optimization using the platform-native languages
CUDA (NVIDIA GPUs), HIP (AMD GPUs), or OpenMP (Intel/AMD/ARM multicore). The
software development cycle ensures production-quality code by featuring unit
testing, automated configuration and installation, Doxygen code documentation,
as well as a Continuous Integration (CI) and Continuous Benchmarking framework.

Ginkgo uses a static template parameter for the value type and a
template parameter for the integer type to allow for compilation in
different precision formats. Standard value type formats supported are
IEEE double precision, IEEE single precision, double complex precision,
and single complex precision. Theoretically, Ginkgo can also be compiled
for any other (arbitrary) precision format, but the support on both the
hardware and the software side is very limited outside the IEEE
standard.

Aside from being compilable for different precision formats, Ginkgo
features the adaptive precision block-Jacobi preconditioner, decoupling
the memory precision from the arithmetic precision, and optimizing the
storage format for the inverted diagonal block
individually. Even though heavily leveraging
advanced multiprecision technology, the numerical considerations of the
adaptive precision block-Jacobi preconditioner are fully automated and
hidden from the user who can employ the functionality as black-box
algorithm without numerical degradation. Building upon the knowledge
gained in the adaptive precision block-Jacobi, Ginkgo is currently
employing the accessor concept to consequently separate the memory
precision from the arithmetic precision,
see~\cref{sec:format_decoupling}.

A orthogonal multiprecision technology that is under consideration for 
integration into Ginkgo is the multiprecision SpMV based on value clustering. 

\subsection{heFFTe}

The Highly-Efficient FFTs for Exascale (heFFTe) library provides fast and
robust multi-dimensional FFT routines for Exascale platforms.  heFFTe leverages
established but \emph{ad hoc} software tools that have traditionally been part
of application codes, but not extracted as independent, supported libraries.
These multidimensional FFTs rely on third-party 1D FFTs, either from FFTW or
from vendor libraries.

FFTs are used in many domain applications--including molecular dynamics,
spectrum estimation, fast convolution and correlation, signal modulation, and
wireless multimedia applications. For example, distributed 3-D FFT is one of
the most important kernels used in molecular dynamics computations, and its
performance can affect an application's scalability on larger machines.
Similarly, the performance of the first principle calculations depends strongly
on the performance of the FFT solver. Specifically, for DOE, we found that more
than a dozen ECP applications use FFT in their codes. However, the current
state-of-the-art FFT libraries are not scalable on large heterogeneous machines
with many nodes, or even on one node with multiple high-performance GPUs (e.g.,
several NVIDIA V100 GPUs). To address these needs, the heFFTe~v0.2 library
release demonstrates very good weak and strong scalability, and a very high
performance that is close to 90\% of the roof-line theoretical peak
performance.  This is achieved through (1)~efficient use of GPUs' high
bandwidth, (2) algorithms to reduce global communications, when possible, and
(3) employment of GPUDirect technologies as well as MPI optimizations. heFFTe provides multi-precision capabilities with support for both single and double precision arithmetic. heFFTe is a C++ library, and the arithmetic used is templated, so that other precisions can be easily added. Current work is on adding mixed-precision capabilities using mixed-preision MPI and compression, as described in Section~\ref{sec:compress}.

\subsection{hypre}
hypre is a software library of high-performance preconditioners and solvers for
the solution of large, sparse linear systems of equations on massively parallel
computers. The hypre library was created with the primary goal of providing
users with advanced parallel preconditioners. The library features parallel
multigrid solvers for both structured and unstructured grid problems. For ease
of use, these solvers are accessed from the application code via hypre's
conceptual linear system interfaces, which allow a variety of natural problem
descriptions and include a structured, a semi-structured and a linear-algebraic
interface. The (semi-)structured interfaces are an alternative to the standard
matrix-based interface, give users a more natural means for describing linear
systems and provide access to structured multigrid solvers, which can take
advantage of the additional information.

\subsection{Kokkos Kernels}

The Kokkos Kernels project primarily focuses on performance-portable kernels for sparse/dense linear algebra and graph kernels. Kokkos Kernels relies on Kokkos programming model for portability.
The focus of sparse linear algebra kernels has been to support the requirements of frameworks such as Trilinos and computational science applications. The sparse linear algebra data structure used is a compressed row storage. Kokkos Kernels provides kernels for sparse matrix-vector multiplication, sparse matrix-matrix multiplication, ILU(k) factorization, Gauss-Seidel preconditioner, triangular solves when the triangular factors arise from direct solvers or incomplete factorizations. All these kernels are templated on the matrix and the vector type allowing multiple precision support from the initial software design. Kokkos Kernels also supports dense linear algebra kernels for team-level BLAS and LAPACK functionality. This allows computational science applications to use BLAS and LAPACK operations in the ``inner-loop'' when programming for accelerators. The BLAS and LAPACK functionality is also templated on the scalar type allowing multiprecision use. Kokkos Kernels also support graph kernels such as distance-1 coloring, distance-2 coloring and triangle counting kernels. 

\subsection{MAGMA}
MAGMA provides LAPACK and a large number of highly optimized dense and
sparse linear algebra (LA) routines for heterogeneous architectures. Besides
LAPACK, other dense LA routines in MAGMA
include BLAS, Batched BLAS and LAPACK, and mixed-precision
factorizations and solvers. A MAGMA Sparse component provides
support for sparse iterative solvers and preconditioners,
a number of sparse matrix formats and conversion routines, SpMV/MM
and auxiliary kernels.

MAGMA addresses the complex challenges of heterogeneous compute
environments by providing hybridized software that combines
the strengths of different algorithms for different hardware components.
MAGMA's LA algorithms target hybrid manycore
systems featuring GPUs specifically and thus enable applications
to fully exploit the power offered by each of the hardware components.
MAGMA provides solvers for linear systems, least squares,
eigenvalue problems, and singular value problems. Designed to be similar
to LAPACK in functionality, data storage, and interface,
the MAGMA library allows scientists to seamlessly port
any linear algebra reliant software components to heterogeneous architectures.
MAGMA allows applications to fully exploit the power of current heterogeneous
systems of multi/many-core CPUs and multi-GPUs to deliver the fastest possible
time to accurate solution within given energy constraints.

MAGMA provides mixed-precision solvers using LU, Cholesky, or QR factorizations. In terms of low precision developments, the latest MAGMA release to-date (v 2.5.3) provides 
an optimized batch HGEMM kernel that outperforms the vendor BLAS for relatively small sizes. It also
provides a mixed-precision linear solver for $Ax=b$ in double-precision, while taking advantage of 
half-precision during the LU factorization. The mixed-precision solver is up to $4\times$ faster 
than a direct FP64 solver, and converges to double precision accuracy if the condition number of 
the matrix $\kappa_\infty(A)$ is up to $10^5$.

\subsection{PETSc}
PETSc is a suite of data structures and routines for the scalable solution of
scientific applications modeled by partial differential equations; TAO is
a scalable library for numerical optimization.  PETSc/TAO can be easily used in application
codes written in C, C++, Fortran, and Python.

PETSc is written in pure C89 (recently extended to support portions of
C999 that are supported by the more recent Microsoft C compilers). The
emphasis has always been on ultimate portability to the Fortran and C
standards. At the same time we have always insured PETSc compilers
completely with C++ as well and that the C compiled version can be used
from C++ compiled code. The largest hassle in this regard has always
been the differences between complex number handling in C and C++
requiring extensive code to handle the differences. PETSc can be  built
only for a single scalar type and precision at a time, for example real
numbers and quad precision. Since C does not offer templates, managing
multiple integrated precision's is difficult. For CPUs, PETSc supports
half-precision (ARM only), single, double, quad (GNU compilers only).
The above applies for CPU based systems. For GPU's, where large
improvements in time to solution are possible with less precision, PETSc
will use its GPU interfaces to allow computing with a selected precision
at runtime on the GPUs. If the numerical values are in, say, double on
the CPU they would be converted to, for example, single when transferred
to the GPU for the computation. Of course, the more desirable case where
the data remains on the GPU will require less conversion, except when
particularly desired, for example, ill-conditioning requires a portion
of the computation to be done with more precision.  

\subsection{PLASMA}

PLASMA (Parallel Linear Algebra Software for Modern
Architectures)~\cite{gullo2009numerical} is a
software package based on modern OpenMP for solving problems
in dense linear algebra. PLASMA provides implementations of
state-of-the-art algorithms using modern task scheduling techniques.
PLASMA provides routines for solving linear systems, least squares
problems, eigenvalue problems, and singular value problems. PLASMA is
based on OpenMP and its data-dependence tracking and task scheduling.
PLASMA library allows scientists to easily port their existing software
components from LAPACK to PLASMA to take advantage of the new multicore
architectures. PLASMA provides LAPACK-style interface for maximum
portability and compatibility. An interface with more efficient data
storage is also provided to achieve performance as close as possible to
the computational peak performance of the machine.

\subsection{SLATE}

SLATE is a distributed, GPU accelerated library for dense linear algebra,
intended as a replacement for ScaLAPACK.
To this end, SLATE provides parallel Basic Linear Algebra Subprograms (BLAS),
norms, linear systems solvers, least square solvers, singular value and
eigenvalue solvers. It is written using modern C++, with ScaLAPACK and LAPACK
compatible wrappers.

SLATE provides mixed-precision solvers using both LU and Cholesky factorization.
The factorization is done in a lower precision, then iterative refinement is
applied to improve the accuracy to a higher precision. The code is templated
on the two precisions; currently single/double and single-complex/double-complex
are supported. Future plans include using half precision and a more robust GMRES
refinement mechanism.

\subsection{STRUMPACK}
STRUMPACK is a distributed, GPU accelerated library for dense and
sparse linear algebra using
rank-structured matrix approximations, including
hierarchically semiseparable (HSS), hierarchically off-diagonal
low rank (HODLR), butterfly, and a non-hierarchical format called
block low rank (BLR).
The baseline sparse STRUMPACK is a multifrontal sparse LU direct solver.
The frontal matrices in the sparse factors can be approximated with
the above rank-structured formats, serving as effective sparse
preconditioners with nearly optimal complexity in flops and memory.
Sparse STRUMPACK relies on ButteryPACK for the HODLR and butterfly formats,
and provides C++ interfaces to the ButteryPACK Fortran library.

STRUMPACK is written using modern C++, with templated datatypes to support
various precisions, including real and complex, single and double
precisions. It can also support half-precision.
Currently, iterative refinement and GMRES
are performed in the same working precision as factorization.

\subsection{SuperLU}
SuperLU is a distributed, GPU accelerated sparse direct solver for
general sparse linear systems, using supernodal techniques in LU factorization and triangular solves.
It uses MPI+OpenMP+CUDA to support various forms of parallelism.
Routines are also provided to equilibrate the system, estimate
the condition number, calculate the relative backward error, and
estimate error bounds for the refined solutions.

SuperLU is written in C and is callable from either C or Fortran program.
The code base uses macros to template the datatypes, so it can support
the mixture of various precisions, including real and complex,
single, double and half precisions. 
Currently, iterative refinement is performed in the
same working precision as factorization. Work is in progress to provide
lower precision factorization coupled with higher precision iterative
refinement.

\subsection{Trilinos}
The Trilinos Project is a premier software framework in scientific computing
for the solution of large-scale, complex multiphysics engineering and
scientific problems. Trilinos is object-oriented and organized into about 60
different packages, each with a specific focus. These packages include linear
and nonlinear solvers, preconditioners (including algebraic multigrid), graph
partitioners, eigensolvers, and optimization algorithms, among other things.
Users are required to install only the subset of packages related to the
problems they are trying to solve.  Trilinos supports MPI+X, where X can be
CUDA, OpenMP, etc. (anything Kokkos supports).

In Trilinos, the scalar type is a template parameter, typically set to IEEE double precision while 
also IEEE single precision is fully supported.
Users can employ preconditioners that are compiled in single precision inside a double 
precision outer solver - however have to account for the numerical effects, i.e., may need a flexible Krylov solver (FCG / FGMRES). A brief discussion of
using mixed precision in the Belos package was given in \cite{Bavier12}.
Other scalar types than single and double may also be used, however, this is not common and not supported in the explicit template instantiation (ETI) build system. 

%% file: formats.tex
\section{IEEE Formats and Format Conversion}
\subsection{Emulator}
The half-precision (fp16) floating-point format, defined in the 2008 revision
of the IEEE standard for floating-point arithmetic, and a more recently
proposed half-precision format bfloat16, are increasingly available in GPUs
and other accelerators. While the support for low precision arithmetic is
mainly motivated by machine learning applications, 
as discussed in earlier sections, general purpose numerical
algorithms can benefit from it too. Since the appropriate 
hardware is not always 
available, and one may wish to experiment with new arithmetics not yet 
implemented in hardware, software simulations of low precision arithmetic are
needed. In \cite{hipr19}, Higham and Pranesh discuss a strategy to
simulate low precision arithmetic using arithmetic
of higher precision, and correctness of such simulations is
explained via rounding error analysis. A MATLAB function 
function \texttt{chop}\footnote{\url{https://github.com/higham/chop}} 
is provided, that can be used to
efficiently simulate fp16, bfloat16, and other low precision arithmetics,
with or without the representation of subnormal numbers and with the options
of round to nearest, directed rounding, stochastic rounding, and random bit
flips in the significand. Interested readers are referred to \cite{hipr19}
for further details.

\subsection{Rounding Error Analysis}

Traditional rounding error analysis in numerical linear algebra leads 
to backward error bounds involving the constant $\gamma_n = nu/(1  -  nu)$, 
for a problem size n and unit roundoff $u$. In
light of large-scale and possibly low-precision computations, 
such bounds can struggle to provide any
useful information. In \cite{hima19a}, Higham and Mary develop 
a new probabilistic rounding 
error analysis that can be applied to a
wide range of algorithms. By using a concentration inequality 
and making probabilistic assumptions
about the rounding errors, they show that in several core linear 
algebra computations $\gamma_n$  can be replaced by a relaxed constant
$\widetilde{\gamma}_n$ proportional to $\sqrt{n \log n} u$ with a 
probability bounded below by a quantity independent of n.
The new constant $\widetilde{\gamma}_n$ grows much more slowly with $n$ than
$\gamma_n$. We refer to \cite{hima19a}, \cite{hima20} for further details. 